\documentclass[12pt,preprint]{aastex}

\def\p    {\phantom{0}}
\def\q    {\phantom{00}}

\def\kms  {km~s$^{-1}$}
\def\kmsperkpc {km~s$^{-1}$~kpc$^{-1}$}
\def\kmskpc    {km~s$^{-1}$~kpc$^{-1}$}
\def\masy {mas~y$^{-1}$}

\def\uas  {\ifmmode {\mu{\rm as}}\else{$\mu$as}\fi}
\def\deg  {\ifmmode {^\circ}\else {$^\circ$}\fi}
\def\porm {\ifmmode {\pm}\else {$\pm$}\fi}
\def\chisqpdf {\ifmmode {\chi^2_{\rm pdf}}\else {$\chi^2_{\rm pdf}$}\fi}
\def\chisq    {\ifmmode {\chi^2}\else {$\chi^2$}\fi}
\def\Msun {M$_\odot$}
\def\HI   {H~{\small I}}

\def\etal {et al.~}
\def\eg   {e.g.,~}
\def\ie   {i.e.,~}

\def\d    {\ifmmode {{\rlap{.}}^\circ}\else {${\rlap{.}}^\circ$}\fi}
\def\s    {\ifmmode {{\rlap{.}}^s}\else {${\rlap{.}}^s$}\fi}
\def\as   {\ifmmode {{\rlap{.}}^{''}}\else {${\rlap{.}}^{''}$}\fi}

\newbox\grsign \setbox\grsign=\hbox{$>$} \newdimen\grdimen \grdimen=\ht\grsign
\newbox\laxbox \newbox\gaxbox
\setbox\gaxbox=\hbox{\raise.5ex\hbox{$>$}\llap
     {\lower.5ex\hbox{$\sim$}}}\ht1=\grdimen\dp1=0pt
\setbox\laxbox=\hbox{\raise.5ex\hbox{$<$}\llap
     {\lower.5ex\hbox{$\sim$}}}\ht2=\grdimen\dp2=0pt
\def\gax{\mathrel{\copy\gaxbox}}
\def\lax{\mathrel{\copy\laxbox}}

\def\pa    {\ifmmode {\psi} \else {$\psi$}\fi}
\def\rPpm  {\ifmmode {r_{\Ro,\To}} \else {$r_{Ro,\To}$}\fi}

\def\vlsr  {\ifmmode {v_{\rm LSR}}\else {$v_{\rm LSR}$}\fi}
\def\vlsrr {\ifmmode {v^r_{\rm LSR}}\else {$v^r_{\rm LSR}$}\fi}
\def\vhelio{\ifmmode {v_{Helio}}\else {$v_{Helio}$}\fi}

\def\ura   {\ifmmode {\mu_\alpha}\else {$\mu_\alpha$}\fi}
\def\udec  {\ifmmode {\mu_\delta}\else {$\mu_\delta$}\fi}

\def\ul    {\ifmmode {\mu_l}\else {$\mu_l$}\fi}
\def\ub    {\ifmmode {\mu_b}\else {$\mu_b$}\fi}

\def\uml   {\ifmmode {v_{gr}}\else {$v_{gr}$}\fi}
\def\umb   {\ifmmode {v_b}\else {$v_b$}\fi}
\def\vsrad {\ifmmode {v_{rad}}\else {$v_{rad}$}\fi}

\def\upl   {\ifmmode {v^p_{gr}}\else {$v^p_{gr}$}\fi}
\def\upb   {\ifmmode {v^p_b}\else {$v^p_b$}\fi}
\def\vprad {\ifmmode {v^p_{rad}}\else {$v^p_{rad}$}\fi}

\def\Vo    {\ifmmode {V^{Std}_\odot}\else {$V^{Std}_\odot$}\fi}
\def\Uo    {\ifmmode {U^{Std}_\odot}\else {$U^{Std}_\odot$}\fi}
\def\Wo    {\ifmmode {W^{Std}_\odot}\else {$W^{Std}_\odot$}\fi}
\def\VH    {\ifmmode {V^H_\odot}\else {$V^H_\odot$}\fi}
\def\UH    {\ifmmode {U^H_\odot}\else {$U^H_\odot$}\fi}
\def\WH    {\ifmmode {W^H_\odot}\else {$W^H_\odot$}\fi}
\def\V     {\ifmmode {V_\odot}\else {$V_\odot$}\fi}
\def\U     {\ifmmode {U_\odot}\else {$U_\odot$}\fi}
\def\W     {\ifmmode {W_\odot}\else {$W_\odot$}\fi}

\def\Vs    {\ifmmode {V_s}\else {$V_s$}\fi}
\def\Us    {\ifmmode {U_s}\else {$U_s$}\fi}
\def\Ws    {\ifmmode {W_s}\else {$W_s$}\fi}

\def\Vsbar {\ifmmode {\overline{V_s}}\else {$\overline{V_s}$}\fi}
\def\Usbar {\ifmmode {\overline{U_s}}\else {$\overline{U_s}$}\fi}
\def\Wsbar {\ifmmode {\overline{W_s}}\else {$\overline{W_s}$}\fi}

\def\aone  {\ifmmode {a_1}\else {$a_1$}\fi}
\def\atwo  {\ifmmode {a_2}\else {$a_2$}\fi}
\def\athr  {\ifmmode {a_3}\else {$a_3$}\fi}

\def\pars  {\ifmmode{\pi_s}\else{$\pi_s$}\fi}

\def\Ts    {\ifmmode{\Theta_s}\else{$\Theta_s$}\fi}
\def\Tdot  {\ifmmode{d\Theta\over dR}\else{$d\Theta\over dR$}\fi}

\def\Rp    {\ifmmode{R_p}\else{$R_p$}\fi}

\def\To    {\ifmmode{\Theta_0}\else{$\Theta_0$}\fi}
\def\Ro    {\ifmmode{R_0}\else{$R_0$}\fi}
\def\Ho    {\ifmmode{H_0}\else{$H_0$}\fi}

\slugcomment{2014 January 28}
\shorttitle{Structure and Kinematics of the Milky Way} 
\shortauthors{Reid \etal}

\begin{document}

\title{TRIGONOMETRIC PARALLAXES OF HIGH MASS STAR FORMING REGIONS: \\
       THE STRUCTURE AND KINEMATICS OF THE MILKY WAY   
       }

\author{M. J. Reid\altaffilmark{1}, K. M. Menten\altaffilmark{2}, 
        A. Brunthaler\altaffilmark{2}, X. W. Zheng\altaffilmark{3}, 
        T. M. Dame\altaffilmark{1},
        Y. Xu\altaffilmark{5}, Y. Wu\altaffilmark{2},
        B. Zhang\altaffilmark{2}, A. Sanna\altaffilmark{2},
        M. Sato\altaffilmark{2}, K. Hachisuka\altaffilmark{6},
        Y. K. Choi\altaffilmark{2}, K. Immer\altaffilmark{2},   
        L. Moscadelli\altaffilmark{4}, K. L. J. Rygl\altaffilmark{7}, 
       \& A. Bartkiewicz\altaffilmark{8}
       }

\altaffiltext{1}{Harvard-Smithsonian Center for
   Astrophysics, 60 Garden Street, Cambridge, MA 02138, USA}
\altaffiltext{2}{Max-Planck-Institut f\"ur Radioastronomie, 
   Auf dem H\"ugel 69, 53121 Bonn, Germany}
\altaffiltext{3}{Department of Astronomy, Nanjing University
   Nanjing 210093, China} 
\altaffiltext{4}{Arcetri Observatory, Firenze, Italy}
\altaffiltext{5}{Purple Mountain Observatory, Chinese Academy of
   Sciences, Nanjing 210008, China}
\altaffiltext{6}{Shanghai Astronomical Observatory, 80 Nandan Rd., Shanghai, China}
\altaffiltext{7}{European Space Agency (ESA-ESTEC), Keplerlaan 1, P.O. Box 299, 2200 AG, Noordwijk, 
The Netherlands}
\altaffiltext{8}{Centre for Astronomy, Faculty of Physics, Astronomy and Informatics, 
   Nicolaus Copernicus University, Grudziadzka 5, 87-100 Torun, Poland}

\begin{abstract}
Over 100 trigonometric parallaxes and proper motions for masers associated with 
young, high-mass stars have been measured with the Bar and Spiral Structure Legacy Survey, 
a Very Long Basline Array key science project, the European VLBI Network, 
and the Japanese VERA project.  These measurements provide strong evidence
for the existence of spiral arms in the Milky Way, accurately locating many arm
segments and yielding spiral pitch angles ranging from about $7^\circ$ to $20^\circ$. 
The widths of spiral arms increase with distance from the Galactic center.
Fitting axially symmetric models of the Milky Way with the 3-dimensional position
and velocity information and conservative priors for the solar and average source peculiar 
motions, we estimate the distance to the Galactic center, $\Ro$, to be $8.34\pm0.16$~kpc, 
a circular rotation speed at the Sun, $\To$, to be $240\pm8$~\kms, 
and a rotation curve that is nearly flat (\ie\ a slope of $-0.2\pm0.4$ \kmsperkpc) 
between Galactocentric radii of $\approx5$ and 16 kpc.  
Assuming a ``universal'' spiral galaxy form for the rotation curve, we estimate 
the thin disk scale length to be $2.44\pm0.16$ kpc.   
With this large data set, the parameters \Ro\ and \To\ are no longer highly 
correlated and are relatively insensitive to different forms of the rotation curve.
If one adopts a theoretically motivated prior that high-mass star forming
regions are in nearly circular Galactic orbits, we estimate a global solar motion 
component in the direction of Galactic rotation, $\V=14.6\pm5.0$ \kms.  
While \To\ and $\V$ are significantly correlated, the sum of these parameters is well 
constrained, $\To+\V = 255.2\pm5.1$ \kms, as is the angular speed of the Sun in its orbit
about the Galactic center, $(\To+\V)/\Ro = 30.57\pm0.43$ \kmskpc.
These parameters improve the accuracy of estimates of the accelerations
of the Sun and the Hulse-Taylor binary pulsar in their Galactic orbits, significantly
reducing the uncertainty in tests of gravitational radiation predicted by 
general relativity.
\end{abstract}


\section{Introduction}

Two major projects to map the spiral structure of the Milky Way are
providing parallaxes and proper motions for water and methanol masers
associated with high-mass star forming regions (HMSFRs) across large 
portions of the Milky Way.  
The Bar and Spiral Structure Legacy (BeSSeL) Survey
\footnote{http://bessel.vlbi-astrometry.org}
 and the Japanese VLBI Exploration of Radio Astrometry (VERA) 
\footnote{http://veraserver.mtk.nao.ac.jp}
have yielded over 100 parallax measurements with accuracies typically about 
$\pm20$ \uas, and some as good as $\pm5$ \uas.  This accuracy exceeds the target 
of the European astrometric satellite mission Gaia, launched in December 2013 and 
scheduled for final results in 2021-2022 \citep{Eyer:13}.  While
Gaia aims to measure $\sim10^9$ stars, far more than practical by 
Very Long Baseline Interferometry (VLBI), Gaia will be limited by
extinction at optical wavelengths and will not be able to freely probe
the Galactic plane.  In contrast, VLBI at radio wavelengths is not affected by 
dust extinction and can yield parallaxes for massive young stars 
that best trace spiral structure in other galaxies, and current parallax
accuracy allows measurements for stars across most of the Milky Way.

Given parallax and proper motion measurements (coupled with source
coordinates and line-of-sight velocities from Doppler shifts of spectral
lines), one has complete phase-space information.  This provides
direct and powerful constraints on the fundamental parameters of
the Galaxy, including the distance to the Galactic center, \Ro, and
the circular orbital speed at the Sun, \To.
Preliminary models of the structure and dynamics of the Galaxy based
on VLBI parallax and proper motions of star forming regions have been
published.  \citet{Reid:09b} fitted results from 16 HMSFRs
and determined $\Ro=8.4\pm0.6$ kpc and $\To=254\pm16$ \kms, assuming
the solar motion in the direction of Galactic rotation, $\V$, is 5 \kms\ 
\citep{Dehnen:98}.  
More recently \citet{Honma:12} analyzed results from a larger sample of 52 sources, 
including both low-mass star forming regions and HMSFRs, and concluded that 
$\Ro=8.05\pm0.45$ kpc and $\To=238\pm14$ \kms, assuming $\V=12$ \kms\ \citep{Schoenrich:10}. 
Several groups have re-modeled maser parallax and proper motion data
\citep{Bovy:09,McMillan:10,Bobylev:10} using different approaches and
focusing on effects of parameter correlations and prior assumptions, 
most notably the values adopted for the solar motion (see \S\ref{sect:priors}
and \S\ref{sect:solar_motion}).

With the much larger number and wider distribution of parallaxes and proper motions 
of HMSFRs now available, we can provide more robust estimates of the fundamental 
Galactic parameters.  In Section \ref{sect:parallaxes}, we present
the combined parallax data sets from the BeSSeL and VERA groups and 
comment on aspects of spiral structure in Section \ref{sect:spiral_structure}.  
We model the combined data set
to obtain better estimates of \Ro\ and \To\ in Section \ref{sect:modeling},
including discussion of priors, different forms of rotation curves, and 
parameter correlations.
Finally, in Section \ref{sect:discussion}, we discuss the solar motion, 
best values for \Ro\ and \To, and some astrophysical implications.

\section{Parallaxes and Proper Motions} \label{sect:parallaxes}

Table~\ref{table:parallaxes} lists the parallaxes and proper motions of 
103 regions of high-mass star formation  measured with VLBI techniques,
using the National Radio Astronomy Observatory's Very Long Baseline Array (VLBA), 
the Japanese VERA project, and the European VLBI Network (EVN).  We have include three 
red supergiants (NML Cyg, S Per, VY CMa) as indicative of HMSFRs, since they are 
high mass stars that have short lifetimes ($<10^7$ yr) and therefore cannot have 
migrated far from their birth locations.  
The locations of these star forming regions in the Galaxy are shown
in Figure~\ref{fig:parallaxes}, superposed on a schematic diagram of the 
Milky Way.   Distance errors are indicated with error bars ($1\sigma$), 
but for many sources the error bars are smaller than the symbols.

\begin{deluxetable}{llrrrrrrll}
\tabletypesize{\tiny} 
\tablecolumns{10} \tablewidth{0pc}
\tablecaption{Parallaxes \& Proper Motions of High-mass Star Forming Regions}
\tablehead {
  \colhead{Source} & \colhead{Alias}& \colhead{R.A.} & \colhead{Dec.} &
  \colhead{Parallax} & \colhead{$\mu_x$} & \colhead{$\mu_y$} &
  \colhead{\vlsr}  & \colhead{Spiral}  & \colhead{Refs.}
\\
  \colhead{}      &\colhead{}      & \colhead{(hh:mm:ss)} & \colhead{(dd:mm:ss)} &
  \colhead{(mas)} & \colhead{(\masy)} & \colhead{(\masy)} &
  \colhead{(\kms)}& \colhead{Arm}   & \colhead{}      
          }
\startdata
G348.70$-$01.04 &               &17:20:04.04 &$-$38:58:30.9 & 0.296$\pm$ 0.026 &  $-$0.73$\pm$ 0.19 &  $-$2.83$\pm$ 0.54 &   $-$7$\pm$  6 &... &1       \\
G351.44$+$00.65 &NGC 6334       &17:20:54.60 &$-$35:45:08.6 & 0.744$\pm$ 0.074 &   0.40$\pm$ 0.51 &  $-$2.24$\pm$ 0.64 &   $-$8$\pm$  3 &Sgr &2         \\
G000.67$-$00.03 &Sgr B2         &17:47:20.00 &$-$28:22:40.0 & 0.129$\pm$ 0.012 &  $-$0.78$\pm$ 0.40 &  $-$4.26$\pm$ 0.40 &   62$\pm$  5 &... &3         \\
G005.88$-$00.39 &               &18:00:30.31 &$-$24:04:04.5 & 0.334$\pm$ 0.020 &   0.18$\pm$ 0.34 &  $-$2.26$\pm$ 0.34 &    9$\pm$  3 &Sct &4           \\
G009.62$+$00.19 &               &18:06:14.66 &$-$20:31:31.7 & 0.194$\pm$ 0.023 &  $-$0.58$\pm$ 0.13 &  $-$2.49$\pm$ 0.29 &    2$\pm$  3 &4$-$k &5       \\
G010.47$+$00.02 &               &18:08:38.23 &$-$19:51:50.3 & 0.117$\pm$ 0.008 &  $-$3.86$\pm$ 0.19 &  $-$6.40$\pm$ 0.14 &   69$\pm$  5 &Con &7         \\
G010.62$-$00.38 &W 31           &18:10:28.55 &$-$19:55:48.6 & 0.202$\pm$ 0.019 &  $-$0.37$\pm$ 0.50 &  $-$0.60$\pm$ 0.25 &   $-$3$\pm$  5 &3$-$k &7     \\
G011.49$-$01.48 &               &18:16:22.13 &$-$19:41:27.2 & 0.800$\pm$ 0.033 &   1.42$\pm$ 0.52 &  $-$0.60$\pm$ 0.65 &   11$\pm$  3 &Sgr &2           \\
G011.91$-$00.61 &               &18:13:58.12 &$-$18:54:20.3 & 0.297$\pm$ 0.031 &   0.66$\pm$ 0.28 &  $-$1.36$\pm$ 0.41 &   37$\pm$  5 &Sct &4           \\
G012.02$-$00.03 &               &18:12:01.84 &$-$18:31:55.8 & 0.106$\pm$ 0.008 &  $-$4.11$\pm$ 0.07 &  $-$7.76$\pm$ 0.27 &  108$\pm$  5 &3$-$k &7       \\
G012.68$-$00.18 &               &18:13:54.75 &$-$18:01:46.6 & 0.416$\pm$ 0.028 &  $-$1.00$\pm$ 0.95 &  $-$2.85$\pm$ 0.95 &   58$\pm$ 10 &Sct &8         \\
G012.80$-$00.20 &               &18:14:14.23 &$-$17:55:40.5 & 0.343$\pm$ 0.037 &  $-$0.60$\pm$ 0.70 &  $-$0.99$\pm$ 0.70 &   34$\pm$  5 &Sct &8         \\
G012.88$+$00.48 &IRAS 18089$-$1732&18:11:51.42 &$-$17:31:29.0 & 0.400$\pm$ 0.040 &   0.15$\pm$ 0.25 &  $-$2.30$\pm$ 0.39 &   31$\pm$  7 &Sct &8,10      \\
G012.90$-$00.24 &               &18:14:34.42 &$-$17:51:51.9 & 0.408$\pm$ 0.025 &   0.19$\pm$ 0.80 &  $-$2.52$\pm$ 0.80 &   36$\pm$ 10 &Sct &8           \\
G012.90$-$00.26 &               &18:14:39.57 &$-$17:52:00.4 & 0.396$\pm$ 0.032 &  $-$0.36$\pm$ 0.80 &  $-$2.22$\pm$ 0.80 &   39$\pm$ 10 &Sct &8         \\
G013.87$+$00.28 &               &18:14:35.83 &$-$16:45:35.9 & 0.254$\pm$ 0.024 &  $-$0.25$\pm$ 2.00 &  $-$2.49$\pm$ 2.00 &   48$\pm$ 10 &Sct &4         \\
G014.33$-$00.64 &               &18:18:54.67 &$-$16:47:50.3 & 0.893$\pm$ 0.101 &   0.95$\pm$ 1.50 &  $-$2.40$\pm$ 1.30 &   22$\pm$  5 &Sgr &9           \\
G014.63$-$00.57 &               &18:19:15.54 &$-$16:29:45.8 & 0.546$\pm$ 0.022 &   0.22$\pm$ 1.20 &  $-$2.07$\pm$ 1.20 &   19$\pm$  5 &Sgr &2           \\
G015.03$-$00.67 &M 17           &18:20:24.81 &$-$16:11:35.3 & 0.505$\pm$ 0.033 &   0.68$\pm$ 0.32 &  $-$1.42$\pm$ 0.33 &   22$\pm$  3 &Sgr &10          \\
G016.58$-$00.05 &               &18:21:09.08 &$-$14:31:48.8 & 0.279$\pm$ 0.023 &  $-$2.52$\pm$ 0.37 &  $-$2.33$\pm$ 0.35 &   60$\pm$  5 &Sct &4         \\
G023.00$-$00.41 &               &18:34:40.20 &$-$09:00:37.0 & 0.218$\pm$ 0.017 &  $-$1.72$\pm$ 0.14 &  $-$4.12$\pm$ 0.33 &   80$\pm$  3 &4$-$k &11      \\
G023.44$-$00.18 &               &18:34:39.19 &$-$08:31:25.4 & 0.170$\pm$ 0.032 &  $-$1.93$\pm$ 0.15 &  $-$4.11$\pm$ 0.13 &   97$\pm$  3 &4$-$k &11      \\
G023.65$-$00.12 &               &18:34:51.59 &$-$08:18:21.4 & 0.313$\pm$ 0.039 &  $-$1.32$\pm$ 0.20 &  $-$2.96$\pm$ 0.20 &   83$\pm$  3 &... &12        \\
G023.70$-$00.19 &               &18:35:12.36 &$-$08:17:39.5 & 0.161$\pm$ 0.024 &  $-$3.17$\pm$ 0.12 &  $-$6.38$\pm$ 0.16 &   73$\pm$  5 &4$-$k &7       \\
G025.70$+$00.04 &               &18:38:03.14 &$-$06:24:15.5 & 0.098$\pm$ 0.029 &  $-$2.89$\pm$ 0.07 &  $-$6.20$\pm$ 0.36 &   93$\pm$  5 &Sct &4         \\
G027.36$-$00.16 &               &18:41:51.06 &$-$05:01:43.4 & 0.125$\pm$ 0.042 &  $-$1.81$\pm$ 0.11 &  $-$4.11$\pm$ 0.27 &   92$\pm$  3 &Sct &10        \\
G028.86$+$00.06 &               &18:43:46.22 &$-$03:35:29.6 & 0.135$\pm$ 0.018 &  $-$4.80$\pm$ 0.30 &  $-$5.90$\pm$ 0.30 &  100$\pm$ 10 &Sct &4         \\
G029.86$-$00.04 &               &18:45:59.57 &$-$02:45:06.7 & 0.161$\pm$ 0.020 &  $-$2.32$\pm$ 0.11 &  $-$5.29$\pm$ 0.16 &  100$\pm$  3 &Sct &6         \\
G029.95$-$00.01 &W 43S          &18:46:03.74 &$-$02:39:22.3 & 0.190$\pm$ 0.019 &  $-$2.30$\pm$ 0.13 &  $-$5.34$\pm$ 0.13 &   98$\pm$  3 &Sct &6         \\
G031.28$+$00.06 &               &18:48:12.39 &$-$01:26:30.7 & 0.234$\pm$ 0.039 &  $-$2.09$\pm$ 0.16 &  $-$4.37$\pm$ 0.21 &  109$\pm$  3 &Sct &6         \\
G031.58$+$00.07 &W 43Main       &18:48:41.68 &$-$01:09:59.0 & 0.204$\pm$ 0.030 &  $-$1.88$\pm$ 0.40 &  $-$4.84$\pm$ 0.40 &   96$\pm$  5 &Sct &6         \\
G032.04$+$00.05 &               &18:49:36.58 &$-$00:45:46.9 & 0.193$\pm$ 0.008 &  $-$2.21$\pm$ 0.40 &  $-$4.80$\pm$ 0.40 &   97$\pm$  5 &Sct &4         \\
G033.64$-$00.22 &               &18:53:32.56 &$+$00:31:39.1 & 0.153$\pm$ 0.017 &  $-$3.18$\pm$ 0.10 &  $-$6.10$\pm$ 0.10 &   60$\pm$  3 &... &1         \\
G034.39$+$00.22 &               &18:53:18.77 &$+$01:24:08.8 & 0.643$\pm$ 0.049 &  $-$0.90$\pm$ 1.00 &  $-$2.75$\pm$ 2.00 &   57$\pm$  5 &Sgr &13        \\
G035.02$+$00.34 &               &18:54:00.67 &$+$02:01:19.2 & 0.430$\pm$ 0.040 &  $-$0.92$\pm$ 0.90 &  $-$3.61$\pm$ 0.90 &   52$\pm$  5 &Sgr &2         \\
G035.19$-$00.74 &               &18:58:13.05 &$+$01:40:35.7 & 0.456$\pm$ 0.045 &  $-$0.18$\pm$ 0.50 &  $-$3.63$\pm$ 0.50 &   30$\pm$  7 &Sgr &14        \\
G035.20$-$01.73 &               &19:01:45.54 &$+$01:13:32.5 & 0.306$\pm$ 0.045 &  $-$0.71$\pm$ 0.21 &  $-$3.61$\pm$ 0.26 &   42$\pm$  3 &Sgr &14        \\
G037.43$+$01.51 &               &18:54:14.35 &$+$04:41:41.7 & 0.532$\pm$ 0.021 &  $-$0.45$\pm$ 0.35 &  $-$3.69$\pm$ 0.39 &   41$\pm$  3 &Sgr &2         \\
G043.16$+$00.01 &W 49N          &19:10:13.41 &$+$09:06:12.8 & 0.090$\pm$ 0.007 &  $-$2.88$\pm$ 0.20 &  $-$5.41$\pm$ 0.20 &   10$\pm$  5 &Per &15        \\
G043.79$-$00.12 &OH 43.8$-$0.1  &19:11:53.99 &$+$09:35:50.3 & 0.166$\pm$ 0.005 &  $-$3.02$\pm$ 0.36 &  $-$6.20$\pm$ 0.36 &   44$\pm$ 10 &Sgr &2         \\
G043.89$-$00.78 &               &19:14:26.39 &$+$09:22:36.5 & 0.121$\pm$ 0.020 &  $-$2.75$\pm$ 0.30 &  $-$6.43$\pm$ 0.30 &   54$\pm$  5 &Sgr &2         \\
G045.07$+$00.13 &               &19:13:22.04 &$+$10:50:53.3 & 0.125$\pm$ 0.005 &  $-$2.98$\pm$ 0.45 &  $-$6.26$\pm$ 0.45 &   59$\pm$  5 &Sgr &2         \\
G045.45$+$00.05 &               &19:14:21.27 &$+$11:09:15.9 & 0.119$\pm$ 0.017 &  $-$2.34$\pm$ 0.38 &  $-$6.00$\pm$ 0.54 &   55$\pm$  7 &Sgr &2         \\
G048.60$+$00.02 &               &19:20:31.18 &$+$13:55:25.2 & 0.093$\pm$ 0.005 &  $-$2.89$\pm$ 0.13 &  $-$5.50$\pm$ 0.13 &   18$\pm$  5 &Per &15        \\
G049.19$-$00.33 &               &19:22:57.77 &$+$14:16:10.0 & 0.189$\pm$ 0.007 &  $-$2.99$\pm$ 0.40 &  $-$5.71$\pm$ 0.40 &   67$\pm$  5 &Sgr &2         \\
G049.48$-$00.36 &W 51 IRS2      &19:23:39.82 &$+$14:31:05.0 & 0.195$\pm$ 0.071 &  $-$2.49$\pm$ 0.14 &  $-$5.51$\pm$ 0.16 &   56$\pm$  3 &Sgr &16        \\
G049.48$-$00.38 &W 51M          &19:23:43.87 &$+$14:30:29.5 & 0.185$\pm$ 0.010 &  $-$2.64$\pm$ 0.20 &  $-$5.11$\pm$ 0.20 &   58$\pm$  4 &Sgr &17        \\
G052.10$+$01.04 &IRAS 19213+1723&19:23:37.32 &$+$17:29:10.5 & 0.251$\pm$ 0.060 &  $-$2.60$\pm$ 2.00 &  $-$6.10$\pm$ 2.00 &   42$\pm$  5 &Sgr &18        \\
G059.78$+$00.06 &               &19:43:11.25 &$+$23:44:03.3 & 0.463$\pm$ 0.020 &  $-$1.65$\pm$ 0.30 &  $-$5.12$\pm$ 0.30 &   25$\pm$  3 &Loc &16        \\
G069.54$-$00.97 &ON 1           &20:10:09.07 &$+$31:31:36.0 & 0.406$\pm$ 0.013 &  $-$3.19$\pm$ 0.40 &  $-$5.22$\pm$ 0.40 &   12$\pm$  5 &Loc &19,20,21  \\
G074.03$-$01.71 &               &20:25:07.11 &$+$34:49:57.6 & 0.629$\pm$ 0.017 &  $-$3.79$\pm$ 1.30 &  $-$4.88$\pm$ 1.50 &    5$\pm$  5 &Loc &21        \\
G075.29$+$01.32 &               &20:16:16.01 &$+$37:35:45.8 & 0.108$\pm$ 0.005 &  $-$2.37$\pm$ 0.11 &  $-$4.48$\pm$ 0.17 &  $-$58$\pm$  5 &Out &22      \\
G075.76$+$00.33 &               &20:21:41.09 &$+$37:25:29.3 & 0.285$\pm$ 0.022 &  $-$3.08$\pm$ 0.60 &  $-$4.56$\pm$ 0.60 &   $-$9$\pm$  9 &Loc &21      \\
G075.78$+$00.34 &ON 2N          &20:21:44.01 &$+$37:26:37.5 & 0.261$\pm$ 0.030 &  $-$2.79$\pm$ 0.55 &  $-$4.66$\pm$ 0.55 &    1$\pm$  5 &Loc &23        \\
G076.38$-$00.61 &               &20:27:25.48 &$+$37:22:48.5 & 0.770$\pm$ 0.053 &  $-$3.73$\pm$ 3.00 &  $-$3.84$\pm$ 3.00 &   $-$2$\pm$  5 &Loc &21      \\
G078.12$+$03.63 &IRAS 20126+4104&20:14:26.07 &$+$41:13:32.7 & 0.610$\pm$ 0.030 &  $-$2.06$\pm$ 0.50 &   0.98$\pm$ 0.50 &   $-$4$\pm$  5 &Loc &24        \\
G078.88$+$00.70 &AFGL 2591      &20:29:24.82 &$+$40:11:19.6 & 0.300$\pm$ 0.024 &  $-$1.20$\pm$ 0.72 &  $-$4.80$\pm$ 0.66 &   $-$6$\pm$  7 &Loc &25      \\
G079.73$+$00.99 &IRAS 20290+4052&20:30:50.67 &$+$41:02:27.5 & 0.737$\pm$ 0.062 &  $-$2.84$\pm$ 0.50 &  $-$4.14$\pm$ 0.70 &   $-$3$\pm$  5 &Loc &25      \\
G079.87$+$01.17 &               &20:30:29.14 &$+$41:15:53.6 & 0.620$\pm$ 0.027 &  $-$3.23$\pm$ 1.30 &  $-$5.19$\pm$ 1.30 &   $-$5$\pm$ 10 &Loc &21      \\
G080.79$-$01.92 &NML Cyg        &20:46:25.54 &$+$40:06:59.4 & 0.620$\pm$ 0.047 &  $-$1.55$\pm$ 0.57 &  $-$4.59$\pm$ 0.57 &   $-$3$\pm$  3 &Loc &26      \\
G080.86$+$00.38 &DR 20          &20:37:00.96 &$+$41:34:55.7 & 0.687$\pm$ 0.038 &  $-$3.29$\pm$ 0.45 &  $-$4.83$\pm$ 0.50 &   $-$3$\pm$  5 &Loc &25      \\
G081.75$+$00.59 &DR 21          &20:39:01.99 &$+$42:24:59.3 & 0.666$\pm$ 0.035 &  $-$2.84$\pm$ 0.45 &  $-$3.80$\pm$ 0.47 &   $-$3$\pm$  3 &Loc &25      \\
G081.87$+$00.78 &W 75N          &20:38:36.43 &$+$42:37:34.8 & 0.772$\pm$ 0.042 &  $-$1.97$\pm$ 0.50 &  $-$4.16$\pm$ 0.51 &    7$\pm$  3 &Loc &25        \\
G090.21$+$02.32 &               &21:02:22.70 &$+$50:03:08.3 & 1.483$\pm$ 0.038 &  $-$0.67$\pm$ 1.56 &  $-$0.90$\pm$ 1.67 &   $-$3$\pm$  5 &Loc &21      \\
G092.67$+$03.07 &               &21:09:21.73 &$+$52:22:37.1 & 0.613$\pm$ 0.020 &  $-$0.69$\pm$ 0.60 &  $-$2.25$\pm$ 0.60 &   $-$5$\pm$ 10 &Loc &21      \\
G094.60$-$01.79 &AFGL 2789      &21:39:58.27 &$+$50:14:21.0 & 0.280$\pm$ 0.030 &  $-$2.30$\pm$ 0.60 &  $-$3.80$\pm$ 0.60 &  $-$46$\pm$  5 &Per &18,28   \\
G095.29$-$00.93 &               &21:39:40.51 &$+$51:20:32.8 & 0.205$\pm$ 0.015 &  $-$2.75$\pm$ 0.20 &  $-$2.75$\pm$ 0.25 &  $-$38$\pm$  5 &Per &28      \\
G097.53$+$03.18 &               &21:32:12.43 &$+$55:53:49.7 & 0.133$\pm$ 0.017 &  $-$2.94$\pm$ 0.29 &  $-$2.48$\pm$ 0.29 &  $-$73$\pm$  5 &Out &27      \\
G100.37$-$03.57 &               &22:16:10.37 &$+$52:21:34.1 & 0.291$\pm$ 0.010 &  $-$3.77$\pm$ 0.60 &  $-$3.12$\pm$ 0.60 &  $-$37$\pm$ 10 &Per &28      \\
G105.41$+$09.87 &               &21:43:06.48 &$+$66:06:55.3 & 1.129$\pm$ 0.063 &  $-$0.21$\pm$ 1.20 &  $-$5.49$\pm$ 1.20 &  $-$10$\pm$  5 &Loc &21      \\
G107.29$+$05.63 &IRAS 22198+6336&22:21:26.73 &$+$63:51:37.9 & 1.288$\pm$ 0.107 &  $-$2.47$\pm$ 1.40 &   0.26$\pm$ 1.40 &  $-$11$\pm$  5 &Loc &29        \\
G108.18$+$05.51 &L 1206         &22:28:51.41 &$+$64:13:41.3 & 1.289$\pm$ 0.153 &   0.27$\pm$ 0.50 &  $-$1.40$\pm$ 1.95 &  $-$11$\pm$  3 &Loc &19        \\
G108.20$+$00.58 &               &22:49:31.48 &$+$59:55:42.0 & 0.229$\pm$ 0.028 &  $-$2.25$\pm$ 0.50 &  $-$1.00$\pm$ 0.50 &  $-$49$\pm$  5 &Per &28      \\
G108.47$-$02.81 &               &23:02:32.08 &$+$56:57:51.4 & 0.309$\pm$ 0.010 &  $-$2.45$\pm$ 1.00 &  $-$3.00$\pm$ 0.70 &  $-$54$\pm$  5 &Per &28      \\
G108.59$+$00.49 &               &22:52:38.30 &$+$60:00:52.0 & 0.398$\pm$ 0.031 &  $-$5.55$\pm$ 0.40 &  $-$3.38$\pm$ 0.40 &  $-$52$\pm$  5 &Per &28      \\
G109.87$+$02.11 &Cep A          &22:56:18.10 &$+$62:01:49.5 & 1.430$\pm$ 0.080 &   0.50$\pm$ 1.50 &  $-$3.70$\pm$ 1.00 &   $-$7$\pm$  5 &Loc &30        \\
G111.23$-$01.23 &               &23:17:20.79 &$+$59:28:47.0 & 0.288$\pm$ 0.044 &  $-$4.28$\pm$ 0.60 &  $-$2.33$\pm$ 0.60 &  $-$53$\pm$ 10 &Per &28      \\
G111.25$-$00.76 &               &23:16:10.36 &$+$59:55:28.5 & 0.294$\pm$ 0.016 &  $-$2.45$\pm$ 0.60 &  $-$2.10$\pm$ 0.60 &  $-$43$\pm$  5 &Per &28      \\
G111.54$+$00.77 &NGC 7538       &23:13:45.36 &$+$61:28:10.6 & 0.378$\pm$ 0.017 &  $-$2.45$\pm$ 0.24 &  $-$2.44$\pm$ 0.25 &  $-$57$\pm$  5 &Per &30      \\
G121.29$+$00.65 &L 1287         &00:36:47.35 &$+$63:29:02.2 & 1.077$\pm$ 0.039 &  $-$0.86$\pm$ 0.76 &  $-$2.29$\pm$ 0.82 &  $-$23$\pm$  5 &Loc &19      \\
G122.01$-$07.08 &IRAS 00420+5530&00:44:58.40 &$+$55:46:47.6 & 0.460$\pm$ 0.020 &  $-$3.70$\pm$ 0.50 &  $-$1.25$\pm$ 0.50 &  $-$50$\pm$  5 &Per &31      \\
G123.06$-$06.30 &NGC 281        &00:52:24.70 &$+$56:33:50.5 & 0.355$\pm$ 0.030 &  $-$2.79$\pm$ 0.62 &  $-$2.14$\pm$ 0.70 &  $-$30$\pm$  5 &Per &32      \\
G123.06$-$06.30 &NGC 281W       &00:52:24.20 &$+$56:33:43.2 & 0.421$\pm$ 0.022 &  $-$2.69$\pm$ 0.31 &  $-$1.77$\pm$ 0.29 &  $-$29$\pm$  3 &Per &19      \\
G133.94$+$01.06 &W 3OH          &02:27:03.82 &$+$61:52:25.2 & 0.512$\pm$ 0.010 &  $-$1.20$\pm$ 0.32 &  $-$0.15$\pm$ 0.32 &  $-$47$\pm$  3 &Per &33,34   \\
G134.62$-$02.19 &S Per          &02:22:51.71 &$+$58:35:11.4 & 0.413$\pm$ 0.017 &  $-$0.49$\pm$ 0.35 &  $-$1.19$\pm$ 0.33 &  $-$39$\pm$  5 &Per &35      \\
G135.27$+$02.79 &WB 89$-$437    &02:43:28.57 &$+$62:57:08.4 & 0.167$\pm$ 0.011 &  $-$1.22$\pm$ 0.30 &   0.46$\pm$ 0.36 &  $-$72$\pm$  3 &Out &36        \\
G160.14$+$03.15 &               &05:01:40.24 &$+$47:07:19.0 & 0.244$\pm$ 0.006 &   0.87$\pm$ 0.35 &  $-$1.32$\pm$ 0.29 &  $-$18$\pm$  5 &... &1         \\
G168.06$+$00.82 &IRAS 05137+3919&05:17:13.74 &$+$39:22:19.9 & 0.130$\pm$ 0.040 &   0.50$\pm$ 0.24 &  $-$0.85$\pm$ 0.17 &  $-$27$\pm$  5 &Out &37,38     \\
G176.51$+$00.20 &               &05:37:52.14 &$+$32:00:03.9 & 1.038$\pm$ 0.021 &   1.84$\pm$ 1.00 &  $-$5.86$\pm$ 1.00 &  $-$17$\pm$  5 &Loc &21        \\
G182.67$-$03.26 &               &05:39:28.42 &$+$24:56:32.1 & 0.149$\pm$ 0.011 &   0.16$\pm$ 0.32 &  $-$0.17$\pm$ 0.32 &   $-$7$\pm$ 10 &Out &37        \\
G183.72$-$03.66 &               &05:40:24.23 &$+$23:50:54.7 & 0.570$\pm$ 0.013 &   0.13$\pm$ 1.20 &  $-$1.40$\pm$ 1.20 &    3$\pm$  5 &Per &28          \\
G188.79$+$01.03 &IRAS 06061+2151&06:09:06.97 &$+$21:50:41.4 & 0.496$\pm$ 0.103 &  $-$0.10$\pm$ 0.50 &  $-$3.91$\pm$ 0.50 &   $-$5$\pm$  5 &Per &39      \\
G188.94$+$00.88 &S 252          &06:08:53.35 &$+$21:38:28.7 & 0.476$\pm$ 0.006 &   0.02$\pm$ 0.30 &  $-$2.02$\pm$ 0.30 &    8$\pm$  5 &Per &18,40       \\
G192.16$-$03.81 &               &05:58:13.53 &$+$16:31:58.9 & 0.660$\pm$ 0.040 &   0.70$\pm$ 0.78 &  $-$1.80$\pm$ 0.86 &    5$\pm$  5 &Per &41          \\
G192.60$-$00.04 &S 255          &06:12:54.02 &$+$17:59:23.3 & 0.628$\pm$ 0.027 &  $-$0.14$\pm$ 0.67 &  $-$0.84$\pm$ 1.80 &    6$\pm$  5 &Per &19        \\
G196.45$-$01.67 &S 269          &06:14:37.08 &$+$13:49:36.7 & 0.189$\pm$ 0.012 &  $-$0.42$\pm$ 0.20 &  $-$0.12$\pm$ 0.20 &   19$\pm$  5 &Out &42        \\
G209.00$-$19.38 &Orion Nebula   &05:35:15.80 &$-$05:23:14.1 & 2.410$\pm$ 0.030 &   3.30$\pm$ 1.50 &   0.10$\pm$ 1.50 &    3$\pm$  5 &Loc &43,44,45      \\
G211.59$+$01.05 &               &06:52:45.32 &$+$01:40:23.1 & 0.228$\pm$ 0.007 &  $-$0.93$\pm$ 0.24 &   0.71$\pm$ 0.26 &   45$\pm$  5 &... &1           \\
G229.57$+$00.15 &               &07:23:01.84 &$-$14:41:32.8 & 0.221$\pm$ 0.014 &  $-$1.34$\pm$ 0.70 &   0.81$\pm$ 0.70 &   47$\pm$ 10 &Per &28          \\
G232.62$+$00.99 &               &07:32:09.78 &$-$16:58:12.8 & 0.596$\pm$ 0.035 &  $-$2.17$\pm$ 0.38 &   2.09$\pm$ 0.60 &   21$\pm$  3 &Loc &40          \\
G236.81$+$01.98 &               &07:44:28.24 &$-$20:08:30.2 & 0.298$\pm$ 0.018 &  $-$3.10$\pm$ 0.63 &   2.12$\pm$ 0.63 &   43$\pm$  7 &Per &28          \\
G239.35$-$05.06 &VY CMa         &07:22:58.33 &$-$25:46:03.1 & 0.855$\pm$ 0.057 &  $-$2.80$\pm$ 0.58 &   2.60$\pm$ 0.58 &   20$\pm$  3 &Loc &46,47       \\
G240.31$+$00.07 &               &07:44:51.92 &$-$24:07:41.5 & 0.212$\pm$ 0.021 &  $-$2.36$\pm$ 0.23 &   2.45$\pm$ 0.30 &   67$\pm$  5 &Per &28          \\
\enddata
\tablecomments {
   Columns 1 and 2 give the Galactic source name/coordinates and an alias, when appropriate.
   Right Ascension and Declination (J2000) are listed in columns 3 and 4.
   Columns 5 through 7 give the parallax and proper motion in the eastward 
   ($\mu_x=\ura \cos{\delta}$) and northward directions ($\mu_y=\udec$).  
   Column 8 lists Local Standard of Rest velocity. Column 9 indicates the
   spiral arm in which it resides, based mostly on association with structure seen
   in $\ell-V$ plots of CO and H~I emission (not using the measured parallaxes);
   starting at the Galactic Center and moving outward, Con=Connecting arm, 3-k=3 kpc arm,
   4-k=4 kpc/Norma arm, Sct=Scutum-Crux-Centaurus arm, Sgr=Sagittarius arm, Loc=Local arm,
   Per=Perseus arm, and Out=Outer arm; a few sources, indicated with ``...'' could not be
   confidently assigned to an arm.   
   Some parameter values listed here were preliminary ones and may be slightly different
   from final values appearing in published papers.  Motion components and their uncertainties 
   are meant to reflect that of the central star that excites the masers, and may be larger
   than formal measurement uncertainties quoted in some papers.  Parallax uncertainties
   for sources with multiple ($N$) maser spots have been adjusted upwards by $\sqrt{N}$,
   if not done so in the original publications. 
   References are   
1:BeSSeL Survey unpublished, 2:\citet{Wu:13}, 3:\citet{Reid:09c}, 4:\citet{Sato:13}, 5:\citet{Sanna:09}, 
6:\citet{Zhang:13b}, 
7:\citet{Sanna:13}, 8:\citet{Immer:13}, 9:\citet{Sato:10a}, 10:\citet{Xu:11}, 
11:\citet{Brunthaler:09}, 12:\citet{Bartkiewicz:08}, 13:\citet{Kurayama:11}, 14:\citet{Zhang:09}, 
15:\citet{Zhang:13}, 16:\citet{Xu:09}, 17:\citet{Sato:10b}, 18:\citet{Oh:10}, 19:\citet{Rygl:10}, 
20:\citet{Nagayama:11}, 21:\citet{Xu:13}, 22:\citet{Sanna:12}, 23:\citet{Ando:11}, 
24:\citet{Moscadelli:11}, 25:\citet{Rygl:12}, 26:\citet{Zhang:12b}, 
27:\citet{Hachi:13}, 28:\citet{Choi:13}, 29:\citet{Hirota:08}, 30:\citet{Moscadelli:09}, 
31:\citet{Moellenbrock:09}, 32:\citet{Sato:08}, 33:\citet{Xu:06}, 34:\citet{Hachi:06}, 
35:\citet{Asaki:10}, 36:\citet{Hachi:09}, 37:\citet{Hachi:13}, 38:\citet{Honma:11}, 
39:\citet{Niinuma:11}, 40:\citet{Reid:09a}, 41:\citet{Shiozaki:11}, 42:\citet{Honma:07}, 
43:\citet{Sandstrom:07},4 4:\citet{Menten:07}, 45:\citet{Kim:08}, 46:\citet{Choi:08}, 47:\citet{Zhang:12a}.   
                }
\label{table:parallaxes}
\end{deluxetable}

Both the proper motion, $\mu_x$ and $\mu_y$, and Local Standard of Rest (LSR) velocity, \vlsr, values and 
their uncertainties are meant to apply to the central star (or stars) that excite 
the masers.  (Note that ``LSR velocities'' are {\it defined} based on the
Standard Solar Motion values of 20 \kms\ toward $18^h$ Right Ascension and 
$30^\circ$ Declination in 1900 coordinates, which translate to Galactic cartesian 
components of $\Uo=10$, $\Vo=15$ and $\Wo=7$ \kms.)
For the \vlsr\ values we adopted methanol maser 
values, when available, or CO emission values from associated giant molecular clouds. 
Since some of the references reporting parallax and proper motion present 
only measurement uncertainty, for these we estimated an additional error term associated with
the uncertainty in transferring the maser motions to that of the central star.
These were added in quadrature with the measurement uncertainties.
For methanol masers, which typically have modest motions of $\lax10$ \kms\  
with respect to the central star, we estimated the additional error term to be 
$\pm5$ \kms\ for \vlsr\ and a corresponding value for the proper motion components
at the measured distance.  While some water masers have expansion motions 
comparable to methanol masers, others display much faster outflow motions.
High velocity outflows are usually associated with water masers that have
spectra rich in features, spread over many tens of \kms.  We, therefore,
evaluated the richness and spread of the water spectra (with respect to
the systemic velocity as indicated by CO emission) and assigned the additional 
error term for $\mu_x$ and $\mu_y$ values between 5 and 20 \kms.

\section{Spiral Structure} \label{sect:spiral_structure}

\begin{figure}
\epsscale{0.85} 
\plotone{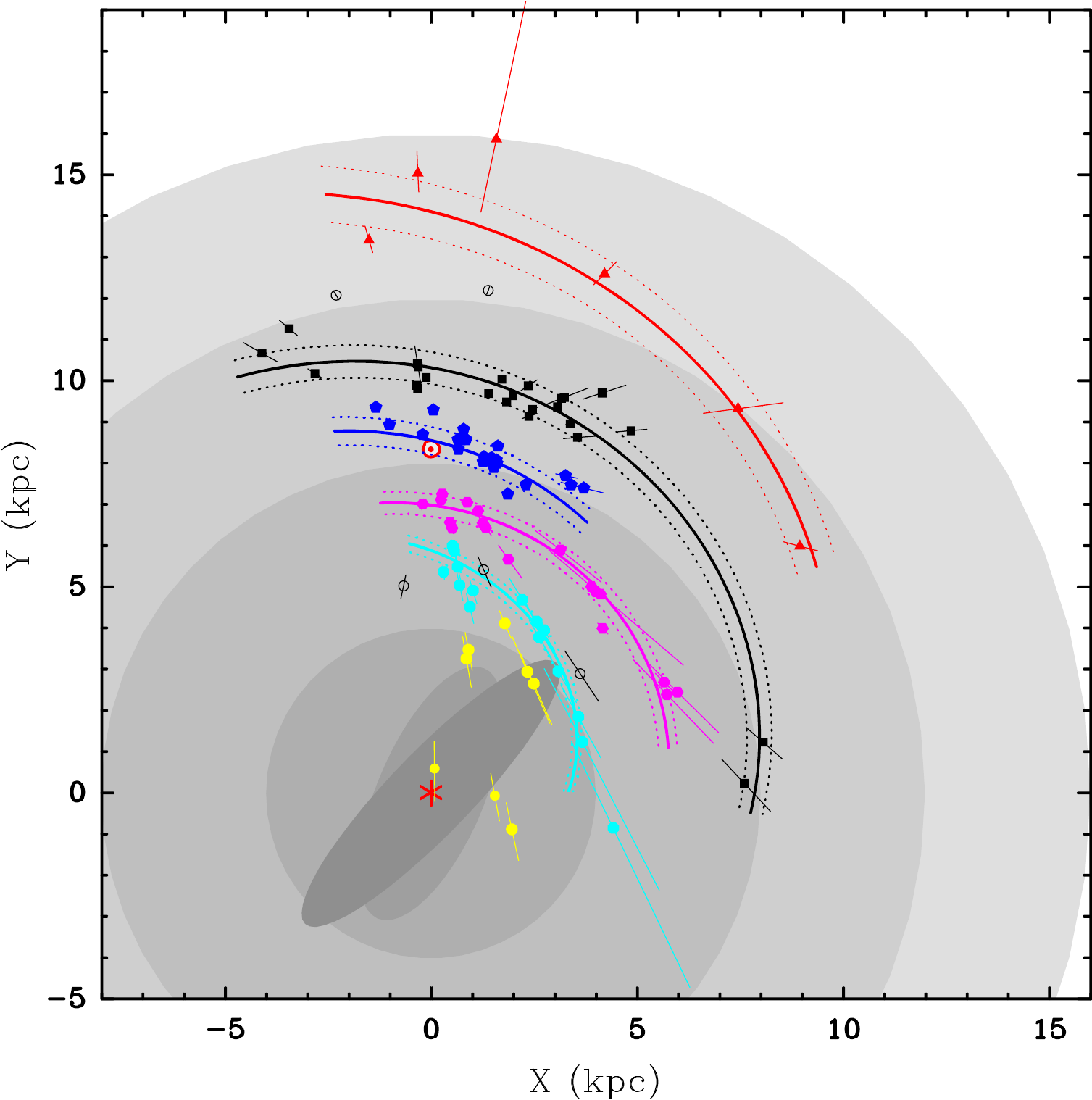}
\caption{\small
Plan view of the Milky Way showing the
locations of high-mass star forming regions (HMSFRs) with 
trigonometric parallaxes measured by the VLBA, VERA, and the EVN.  
The Galactic center ({\it red asterisk}) is at (0,0) and the Sun 
({\it red Sun symbol}) is at (0,8.34).
HMSFRs were assigned to spiral arms based primarily on association with 
structure seen in $\ell-V$ plots of CO and \HI\ emission (and not based on the 
measured parallaxes): Inner Galaxy sources, {\it yellow dots}; Scutum arm, 
{\it cyan octagons};
Sagittarius arm, {\it magenta hexagons}; Local arm, {\it blue pentagons}; 
Perseus arm, {\it black squares}; Outer arm, {\it red triangles}.
Open circles indicate sources for which arm assignment was unclear.
Distance error bars are indicated, but many are smaller than the symbols.
The background grey disks provide scale, with radii corresponding in round numbers
to the Galactic bar region ($\approx4$ kpc), the solar circle ($\approx8$ kpc), 
co-rotation of the spiral pattern and Galactic orbits ($\approx12$ kpc), 
and the end of major star formation ($\approx16$ kpc).
The short COBE ``boxy-bar'' and the ``long'' bar \citep{Blitz:91,Hammersley:00,Benjamin:08}
are indicated with shaded ellipses.
The {\it solid} curved lines trace the centers (and {\it dotted} lines the $1\sigma$ widths) 
of the spiral arms from the log-periodic spiral fitting (see \S3 and Table 2).
For this view of the Milky Way from the north Galactic pole, Galactic rotation is clockwise.}
\label{fig:parallaxes}
\end{figure}

Spiral arms in the Milky Way have long been recognized as presenting coherent arcs and 
loops in Galactic longitude--velocity ($\ell-V$) plots of atomic and molecular emissions.
However, transforming velocity to distance (\ie kinematic distances) has
been problematic, owing to near-far distance ambiguities in the first and
fourth Galactic quadrants and significant distance errors owing to large peculiar 
motions for some arm material (see, \eg\ \citet{Xu:06,Reid:09b}).  
While one cannot accurately place spiral arms on a plan view of the Milky Way 
from $\ell-V$ plots, one can in most cases unambiguously assign HMSFRs 
to spiral arms by association with CO and \HI\ emission features.  
We have done this for the vast majority of the HMSFRs for which parallax and proper 
motions have been measured \citep{Hachi:13,Choi:13,Zhang:13,Xu:13,Wu:13,Sato:13,Sanna:13}, 
as indicated in Table \ref{table:parallaxes} and Figure \ref{fig:parallaxes}.  
This avoids using the measured distances (parallaxes) and subjective judgment 
based on spatial location for arm assignments.

There are two avenues for checking that the arm assignments are reliable.   
Firstly, and most straightforwardly, looking at a plan view of the Milky Way 
(see Fig.~\ref{fig:parallaxes}) on which star forming regions with parallax 
distances are located, one can see that the pattern of sources for any given 
arm traces a continuous arc that resembles a spiral arm in external galaxies.  
Also, there are clear inter-arm regions with 
few, if any, HMSFRs between the Outer, Perseus, Local, Sagittarius, and Scutum arms.   
However, as one looks to the inner Galaxy, the current parallax data are not adequate 
to clearly separate arms, presuming significant separations even exist.    

Secondly, once sources are assigned to arms based on $\ell-V$ information, one
can then attempt to fit their radial and azimuthal locations to log-periodic spiral 
forms using measured distances.  In the papers cited above, we fitted spiral patterns
to arm segments, adopting a log-periodic spiral defined by   
$$\ln{(R/R_{ref})} = -(\beta - \beta_{ref}) \tan{\pa}~~,$$
where $R$ is the Galactocentric radius at a Galactocentric azimuth $\beta$ 
(defined as 0 toward the Sun and increasing with Galactic longitude) for an arm with 
a radius $R_{ref}$ at reference azimuth $\beta_{ref}$ and pitch angle $\pa$.  
We fitted a straight line to ($x,y$)=($\beta,\ln{(R/R_{ref})}$) using a Bayesian Markov 
chain Monte Carlo (McMC) procedure to estimate the parameters $R_{ref}$ and $\pa$.
(The reference azimuth, $\beta_{ref}$, was arbitrarily set near the midpoint of the 
azimuth values for the sources in an arm).  
We minimized the ``distance'' perpendicular to the fitted straight 
line by rotating ($x,y$) through the angle $\pa$ to ($x_r,y_r$), \ie\
$$x_r = x~\cos{\pa} + y~\sin{\pa};~~~y_r = y~\cos{\pa} - x~\sin{\pa}~~,$$
such that the best-fitting line lay in the $x_r$ axis.

Uncertainties in the source parallax ``map'' into both coordinates and 
were estimated numerically by randomly drawing trial parallax values (consistent with
the measured values and uncertainties) and calculating the root-mean-squares for trial
$\ln{(R/R_{ref})}$ and $\beta$ values.  
The locations of the HMSFRs deviated from fitted spirals by more 
than could be explained by parallax uncertainties.  This is expected for spiral 
arms with intrinsic widths of several hundred parsecs.
In order to allow for (and estimate) the scatter in 
location expected from the width of the spiral arm, before calculating trial
$\ln{(R/R_{ref})}$ values, we added random scatter to the trial $R$ values via
$R \leftarrow R + g a_w \cos{\pa}$, where $g$ is a random number drawn from
a Gaussian distribution with zero mean and unity standard deviation and $a_w$
is an arm-width parameter, adjusted to give a post-fit $\chi^2_\nu$ near unity.  
The uncertainties in ($\beta,\ln{(R/R_{ref})}$) were then rotated by angle $\pa$ to 
match the data.

The sum of the squares of the residuals divided by their uncertainties in the $y_r$ direction 
were minimized.  Since preliminary estimates of $\pa$ affect these quantities,
we iterated the fitting to convergence.  Final parameter values were
estimated from marginalized posteriori probability density distribution functions (PDFs) 
for each parameter based on McMC trials that were accepted 
or rejected following the Metropolis-Hastings algorithm; the values 
reported in Table \ref{table:pitchangles} assume $\Ro=8.34$ kpc 
(see \S\ref{sect:modeling}).   Based on the fitted parameter values, we plot
the trace of the centers and $1\sigma$ widths of each arm on Fig. \ref{fig:parallaxes}.

\begin{figure}
\epsscale{1.0} 
\plotone{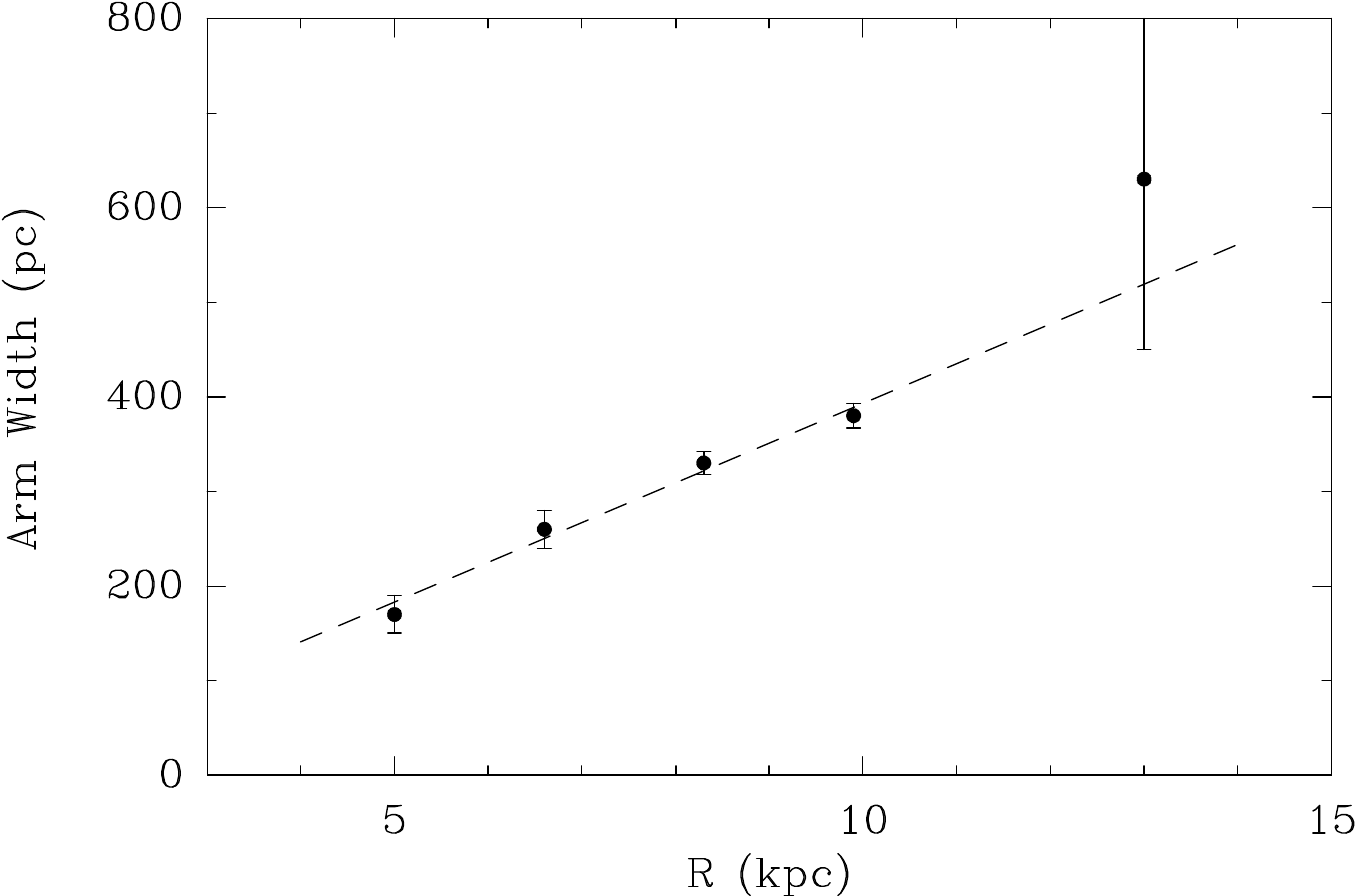}
\caption{\small
  Spiral arm width increasing with Galactocentric radius.  The {\it dashed line} is
a variance-weighted fit with a slope of 42 pc kpc$^{-1}$.  See Table \ref{table:pitchangles}
for details.
        }
\label{fig:armwidths}
\end{figure}

The intrinsic widths of the spiral arms, estimated from the $a_w$ parameters, show an interesting 
pattern in Fig.~\ref{fig:armwidths}.   The estimated arm widths increase nearly linearly with 
Galactocentric radius at a rate of 42 pc kpc$^{-1}$ between radii of 5 to 13 kpc.
Spiral pitch angles vary between $7^\circ$ and $20^\circ$ as listed in 
Table \ref{table:pitchangles}.  The significant range of 
pitch angles among arms suggests that no single value applies to all arms and, possibly, 
cannot be applied to the full length of an arm as it winds around the Galaxy
\citep{Savchenko:13}.  
However, these pitch angles are characteristic of spiral galaxies of Sb to Sc class 
\citep{Kennicutt:81}, further supporting the identification of $\ell-V$ tracks as 
spiral arms for the Milky Way.

\begin{deluxetable}{lrlrrr}
\tablecolumns{6} \tablewidth{0pc} 
\tablecaption{Spiral Arm Characteristics}
\tablehead {
 \colhead{Arm} &\colhead{N} &\colhead{$\beta_{ref}$ ($\beta$ range)} &\colhead{$R_{ref}$} &\colhead{Width} &\colhead{$\psi$} \\
 \colhead{}    &\colhead{}  &\colhead{(deg)}         &\colhead{(kpc)}     &\colhead{(kpc)} &\colhead{(deg)} 
           }
\startdata
Scutum         &17          &27.6 ($+3\rightarrow101$) &$5.0\pm0.1$         & $0.17\pm0.02$ &$19.8\pm2.6$ \\
Sagittarius    &18          &25.6 ($-2\rightarrow68$)  &$6.6\pm0.1$         & $0.26\pm0.02$ &$6.9\pm1.6$  \\
Local          &25          &\p8.9 ($-8\rightarrow27$) &$8.4\pm0.1$         & $0.33\pm0.01$ &$12.8\pm2.7$ \\
Perseus        &24          &14.2 ($-21\rightarrow88$) &$9.9\pm0.1$         & $0.38\pm0.01$ &$9.4\pm1.4$  \\
Outer          &6           &18.6 ($-6\rightarrow56$)  &$13.0\pm0.3$        & $0.63\pm0.18$ &$13.8\pm3.3$ \\
\enddata
\tablecomments{\footnotesize
~Spiral arm data from fitting a section of a log-periodic spiral for the arms listed in column 1.
See the primary papers for more information on each arm:
Scutum arm \citep{Sato:13}, Sagittarius arm \citep{Wu:13}, 
Local arm \citep{Xu:13}, Perseus arm \citep{Choi:13,Zhang:13}, Outer arm \citep{Hachi:13}.
Small differences between parameter values in these papers and here reflect small differences 
between preliminary and final parallax values and the
adopted value for \Ro; here we use $\Ro\equiv8.34$ kpc.
For the Local arm, the pitch angle fit here used only HMSFRs.
Column 2 lists the number of HMSFRs with parallax measurements used in the fits.  Columns 3 and 4
give the reference Galactocentric azimuth, an arbitrary value assigned near the center of the range of source
azimuths (given in parentheses), and the fitted radius at that azimuth.  Column 5 is an estimate of the 
intrinsic arm width, based on the magnitude of ``astrophysical noise'' added to the measurement
uncertainty to achieve a $\chi_\nu^2$ per degree of freedom near unity.
Column 6 is the spiral arm pitch angle, a measure of how tightly wound the spiral is.
               }
\label{table:pitchangles}
\end{deluxetable}

The HMSFRs with measured parallaxes are clearly tracing the 
major spiral arms of the Milky Way (see Fig.~\ref{fig:parallaxes}), and  
details of the locations and properties of the individual arms can be
found in the primary references 
\citep{Hachi:13,Choi:13,Zhang:13,Xu:13,Wu:13,Sato:13,Sanna:13}.
Interestingly, some surprising results are already evident. 
We are finding that the Perseus arm, thought to be one of the major spiral arms of the
Milky Way, has little massive star formation over a 6 kpc-long arc between Galactic 
longitudes of $50^\circ$ and $80^\circ$ \citep{Choi:13,Zhang:13}.  
On the other hand, the Local (Orion) arm, often called a ``spur'' and considered a minor 
structure \citep{Blaauw:85}, has comparable massive star formation to its adjacent 
Sagittarius and Perseus arms \citep{Xu:13}.

\section{Modeling the Galaxy} \label{sect:modeling}

Given measurements of position, parallax, proper motion and Doppler shift,
one has complete three-dimensional location and velocity vectors relative 
to the Sun.  One can then construct a model of the Milky Way and adjust the 
model parameters to best match the data.  As in \citet{Reid:09b}, 
we model the Milky Way as a disk rotating with speed 
$\Theta(R)=\To+\Tdot~(R-\Ro)$, 
where \Ro\ is the distance from the Sun to the Galactic center and \To\ is
the circular rotation speed at this distance.  We then evaluate the
effects of different forms for the rotation curve.
Since all measured motions are relative to the Sun, we need to model
the peculiar (non-circular) motion of the Sun, parameterized by 
\U\ toward the Galactic center, \V\ in the direction of Galactic rotation, and 
\W\ towards the north Galactic pole (NGP).  
Table~\ref{table:model} summarizes these and other parameters.

\begin{deluxetable}{ll}
\tablecolumns{2} \tablewidth{0pc} 
\tablecaption{Galaxy Model Parameter Definitions}
\tablehead {
  \colhead{Parameter} & \colhead{Definition} 
            }
\startdata
\Ro   & Distance of Sun from GC \\
\To   & Rotation Speed of Galaxy at \Ro\\
\Tdot & Derivative of $\Theta$ with $R$: $\Theta(R)=\To+\Tdot~(R-\Ro)$\\
\\
\U    & Solar motion toward GC  \\
\V    & Solar motion in direction of Galactic rotation\\
\W    & Solar motion toward NGP \\
\\
\Usbar   & Average source peculiar motion toward GC  \\
\Vsbar   & Average source peculiar motion in direction of Galactic rotation\\
\Wsbar   & Average source peculiar motion toward NGP \\
\enddata
\tablecomments {\footnotesize
GC is the Galactic Center and NGP is the North Galactic Pole.  
The average source peculiar motions (\Usbar,\Vsbar,\Wsbar) are defined at the 
location of the source and are rotated with respect to the solar motion (\U,\V,\W) 
by the Galactocentric azimuth, $\beta$, of the source 
(see Figure~8 of \citet{Reid:09b}).
We solve for the magnitude of each velocity component, but the orientation of 
the vector for each source depends on location in the Galaxy.
               } 
\label{table:model}
\end{deluxetable}

For each source, we treated the 3-dimensional velocity components (two components of 
proper motion, $\mu_x$ and $\mu_y$, and the heliocentric Doppler velocity, \vhelio, as data
to be compared to a model.  The source coordinates ($\ell,b$) and parallax distance 
($1/\pars$) were treated as independent variables.  
This approach is slightly different than in \citet{Reid:09b}, where
the parallaxes were also treated as data in the least-squares fitting.  While that
approach adds some extra information (\eg\ for sources near the Galactic
tangent points, distance is very sensitive to Doppler velocity, but 
not vice versa), it brings correlated data into the fitting, which will lead to slightly
underestimated parameter uncertainties.   We tested the inclusion versus exclusion of 
parallax with simulated data sets and found little difference and no bias between the 
methods.  However, in order to avoid the need to adjust formal parameter uncertainties, 
as well as subtle issues associated with resolving the near/far distance ambiguities
for sources in the first and fourth Galactic quadrants,
we used the more conservative ``velocity-only'' fitting as done, for example, by
others \citep{Bovy:09,McMillan:10,Bobylev:10,Honma:12}.

\subsection{Bayesian fitting}\label{sect:Bayesian}

We adjusted the Galactic parameters so as to best match the data to the 
spatial-kinematic model using a Bayesian fitting approach.  
The posteriori PDFs of the parameters
were estimated with Markov chain Monte Carlo (McMC) trials that
were accepted or rejected by the Metropolis--Hastings algorithm.  
While a simple axi-symmetric model for the Galaxy may be a reasonable
approximation for the majority of sources, a significant minority of outliers
are expected for a variety of well known reasons.  For example, the gravitational
potential of the Galactic bar (or bars), which extend 3 to 4 kpc from the Galactic center 
\citep{Liszt:80,Blitz:91,Hammersley:00,Benjamin:05} is expected to induce large
non-circular motions for sources in its vicinity.  Indeed, some of these
sources show large peculiar motions, although based on a nearly flat rotation 
curve extrapolated inward from measurements outside this region \citep{Sanna:13}.
Therefore, we removed the eight sources within 4 kpc of the Galactic center
(\ie excluding G000.67$-$00.03, G009.62$+$00.19, G010.47$+$00.02,
G010.62$-$00.38, G012.02$-$00.03, G023.43$-$00.18, G023.70$-$00.19, 
G027.36$-$00.16) before model fitting.  

In the Galaxy's spiral arms, super-bubbles created by multiple supernovae 
can accelerate molecular clouds to $\approx20$ \kms\ \citep{Sato:08}.  
It is probably not possible, prior to fitting, to determine which sources have 
been thus affected and are likely kinematically anomalous.  Therefore,
we initially used an ``outlier-tolerant'' Bayesian fitting scheme described 
by \citet{Sivia:06} as a ``conservative formulation,'' which minimizes the
effects of deviant points on estimates of the fitted parameters.   
For this approach, one maximizes 
$$\sum_{i=1}^N~\sum_{j=1}^3~{\ln\bigl(~(1-e^{-R_{i,j}^2/2})/R_{i,j}^2~\bigr)}~~,$$ 
where the weighted residual $R_{i,j} = (v_{i,j}~-~m_{i,j})/w_{i,j}$
(\ie\ the data ($v$) minus model ($m$) divided by the uncertainty ($w$) 
of the $i^{th}$ of $N$ sources and $j^{th}$ velocity component).   
For large residuals, this formulation assigns a $1/R^2$ probability, 
compared to a Gaussian probability of $e^{-R^2/2}$ which vanishes rapidly.  
Thus, for example, a $5\sigma$ outlier has a reasonable (4\%) probability with 
the outlier-tolerant approach, compared to $\approx10^{-6}$ probability for 
Gaussian errors in the least-squares method, and will not be given excessive 
weight when adjusting parameters.  Once the outliers were identified and removed, 
we assumed Gaussian data uncertainties and fitted data by maximizing 
$$\sum_{i=1}^N~\sum_{j=1}^3~{-R_{i,j}^2/2}~~,$$ 
essentially least-squares fitting. 

Our choice of weights ($w$) for the data in the model fitting process 
was discussed in detail in \citet{Reid:09b}.  We include both measurement uncertainty 
and the effects of random (Virial) motions of a massive young star (with maser emission) 
with respect to the average motion of the much larger and more massive HMSFR
when weighting the differences between observed and modeled components of motion.  
Specifically the proper motion and Doppler velocity weights were given by 
$w(\mu) = \sqrt{\sigma^2_\mu + \sigma^2_{Vir}/d^2_s}$ and
$w(\vhelio) = \sqrt{\sigma^2_v + \sigma^2_{Vir}}$, where $\sigma^2_{Vir}$ is the expected
(1-dimensional) Virial dispersion for stars in a high mass star forming region (HMSFR).
We adopted $\sigma_{Vir}=5$ \kms, appropriate for HMSFRs with $\sim10^4$ \Msun
within a radius of $\sim1$ pc, and did not adjust this value. 
As will be seen in \S\ref{sect:Bayesian}, the vast majority of the velocity
data can be fit with a $\chi^2_\nu$ near unity with these weights.
Note that we were fairly conservative when assigning motion uncertainties 
for individual stars based on the maser data (see \S\ref{sect:parallaxes}),
and this may result in a slightly low $\sigma_{Vir}$ value in order to
achieve unity $\chi^2_\nu$ fits.

\subsection{Priors}\label{sect:priors}

In order to model the observations, one needs prior constraints on the non-circular
motion of our measurement ``platform'' (\ie\ the solar motion parameterized by
\U, \V, \W) and/or the average peculiar motion of the sources being 
measured (parameterized by \Usbar, \Vsbar, \Wsbar).  
Allowing for a non-zero average source peculiar motion can be thought of as a first
approximation of the kinematic effects of spiral structure.  
In \citet{Reid:09b}, we assumed the solar motion determined by 
\citet{Dehnen:98} based on Hipparcos measurements and concluded that HMSFRs 
lagged circular orbital speeds by 15 \kms\ (\ie $\Vsbar=-15$ \kms).  
The observed orbital lag ($\Vsbar<0$) is insensitive 
to the value adopted for \To, but it is strongly correlated with the adopted 
solar motion component, $\V$ \citep{Reid:09b,Honma:12}.  
Recently, the value of the solar motion component in the direction of 
Galactic rotation (\V) has become controversial.  Motivated in part by the
large \Vsbar\ lag in \citet{Reid:09b}, \citet{Schoenrich:10} re-evaluated the 
standard ``asymmetric-drift'' approach used by \citet{Dehnen:98} and 
concluded that it was biased by coupled metallicity/orbital-eccentricity 
effects.  They suggested new solar motion values; specifically they argued for 
a substantial increase for \V\ from 5 to 12 \kms.  This change would decrease 
the average orbital lag of HMSFRs (\Vsbar) by $\approx7$ \kms\ to a more
theoretically appealing value near 8 \kms. 

Based on the first year of data from the Apache Point Observatory 
Galactic Evolution Experiment (APOGEE), \citet{Bovy:12} argue that 
the Sun's motion relative to a circular orbit in the Galaxy 
(ie, a ``rotational standard of rest'') is 26 \kms\ in the direction of 
Galactic rotation, suggesting that the entire Solar Neighborhood, 
which defines the local standard of rest (LSR), leads a circular orbit by 14 \kms.  
Taking into account these developments, we considered a conservative prior of 
$\V = 15\pm10$ \kms, that encompasses the values of \V\ from 5 to 26 \kms\ within 
approximately the $\pm1\sigma$ range.

One could argue on theoretical grounds that HMSFRs should, on average, 
lag circular orbits by only a few \kms\ \citep{McMillan:10}.   
We observe masers in HMSFRs that are very young  
and the gas out of which their exciting stars formed could have responded to 
magnetic shocks when entering spiral arms, leading to departures from circular 
speeds by $\lax10$ \kms\ \citep{Roberts:70}, apportioned between components 
counter to rotation and toward the Galactic Center.  In addition, radial 
pressure gradients can also reduce orbital speeds of gas slightly \citep{Burkert:10}, 
contributing to a small lag of $\approx1$ \kms.   Allowing for such effects, 
we consider priors for \Usbar\ of $3\pm10$ \kms\ and \Vsbar\ of $-3\pm10$ \kms\ 
as reasonable and conservative.

Given the current uncertainty in a) the value for the circular (\V) component of 
solar motion and b) the magnitude of the average peculiar motions of HMSFRs, 
we tried four sets of priors when fitting the data:
\begin{itemize}
\item[Set-A)]{} Adopting a loose prior for the \V\ component of solar motion,
$\U = 11.1\pm1.2$, $\V = 15\pm10$, $\W = 7.2\pm1.1$ \kms, and for the
average peculiar motions for HMSFRs of $\Usbar = 3\pm10$ and 
$\Vsbar =-3\pm10$ \kms.
\item[Set-B)]{} Using no priors for the average peculiar motions of HMSFRs, but 
tighter priors for the solar motion of $\U = 11.1\pm1.2$, $\V = 12.2\pm2.1$, 
$\W = 7.2\pm1.1$ \kms\ from \citet{Schoenrich:10}.
\item[Set-C)]{} Using no priors for the solar motion, but tighter priors on the 
average peculiar motions of HMSFRs of $\Usbar = 3\pm5$ and $\Vsbar =-3\pm5$ \kms.
\item[Set-D)]{} Using essentially no priors for either the solar or 
average peculiar motions of HMSFRs, but bounding the \V\ and \Vsbar\ parameters
with equal probability within $\pm20$ \kms\ of the Set-A initial values and zero
probability outside that range.
\end{itemize}

\subsection{Models A1--A4}

Using the 95 sources with Galactocentric radii greater than 4 kpc\footnote{
Removing sources for which $R<4$ kpc: 
G$000.67-00.03$, G$009.62+00.19$, G$010.47+00.02$, G$010.62-00.38$, 
G$012.02-00.03$, G$023.43-00.18$, G$023.70-00.19$, G$027.36-00.16$}, 
the outlier-tolerant Bayesian fitting approach, and the Set-A priors as described 
above, we obtained the parameter estimates listed in Table \ref{table:fits} under fit A1.  
As expected for a sample with some outliers (see discussion in \S\ref{sect:Bayesian}), 
we found a $\chisq=562.6$, greatly exceeded the 277 degrees of freedom, 
owing to a number of sources with large residuals.   
  
We iteratively removed the sources with the largest residuals.  
Using the outlier-tolerant Bayesian fitting approach (see \S\ref{sect:Bayesian}) 
minimizes potential bias, based on assumed ``correct'' parameter values, when editing data.  
However, to further guard against any residual bias, we first 
removed sources with $>6\sigma$ residuals, followed by re-fitting and removal of 
those with $>4\sigma$ residuals, and finally re-fitting and removal of those with 
$>3\sigma$ residuals (fits A2, A3 \& A4, not listed here).  In total, 15 
sources\footnote{Removing outlying sources: 
G012.68$-$00.18, G016.58$-$00.05, G023.65$-$00.12, G025.70$+$00.04, G028.86$+$00.06, 
G029.95$-$00.01, G031.28$+$00.06, G033.64$-$00.22, G034.39$+$00.22, G078.12$+$03.63, 
G108.59$+$00.49, G111.54+0.77, G122.01$-$07.08, G133.94+01.06, G176.51$+$00.20} 
were removed. 

\begin{deluxetable}{lccccc}
\tablecolumns{6} \tablewidth{0pc} 
\tablecaption{Bayesian Fitting Results}
\tablehead {
  \colhead{} & \colhead{A1} & \colhead{A5} &  \colhead{B1}  & 
  \colhead{C1} & \colhead{D1}   
           }
\startdata
Parameter Estimates \\
\Ro~(kpc)    &$8.15\pm0.25$   &$8.34\pm0.16$  &$8.33\pm0.16$  &$8.30\pm0.19$  &~$8.29\pm0.21$ \\
\To~(\kms)   &~$238\pm11\q$   &$240\pm8\q$    &$243\pm6\q$    &$239\pm8\q$    &$238\pm15$     \\
\Tdot~(\kmskpc)&$-0.1\pm0.7\q$&$-0.2\pm0.4\q$ &$-0.2\pm0.4\q$ &$-0.1\pm0.4\q$ &$-0.1\pm0.4$   \\
\\
\U~(\kms)    &$10.4\pm1.8$    &$10.7\pm1.8$   &$10.7\pm1.8$   &$ 9.9\pm3.0$   &$ 9.6\pm3.9$   \\
\V~(\kms)    &$15.1\pm7.3$    &$15.6\pm6.8$   &$12.2\pm2.0$   &$14.6\pm5.0$   &$16.1\pm13.5$  \\
\W~(\kms)    &\p$8.2\pm1.2$   &$\p8.9\pm0.9$  &$\p8.7\pm0.9$  &$\p9.3\pm1.0$  &$\p9.3\pm1.0$  \\
\\
\Usbar~(\kms)&$\p3.7\pm2.4$   &$\p2.9\pm2.1$  &$\p2.9\pm2.1$  &$\p2.2\pm3.0$  &$\p1.6\pm3.9$  \\
\Vsbar~(\kms)&$-2.4\pm7.4$    &$-1.6\pm6.8$   &$-5.0\pm2.1  $ &$-2.4\pm5.0$   &$-1.2\pm13.6$  \\
\\
Fit Statistics \\
$\chisq$         & 562.6      & 224.9         & 225.1         & 224.7         & 224.1         \\
$N_{dof}$        & 277        & 232           & 232           & 232           & 232           \\
$N_{sources}$    & 95         & 80            & 80            &  80           &  80           \\
$r_{\Ro,\To}$    &0.61        &0.46           &0.74           &0.66           &0.44           \\
\enddata
\tablecomments{\footnotesize
~Fit A1 used the 95 sources in Table~\ref{table:parallaxes} for which
Galactocentric radii exceeded 4 kpc, an ``outlier tolerant''  
probability distribution function for the residuals (see \ref{sect:Bayesian}),
and Set-A priors: Gaussian solar motion priors of $\U=11.1\pm2.0$, $\V=15\pm10$, 
$\W=7.2\pm2.0$ \kms\ and average source peculiar motion priors of 
$\Usbar=3\pm10$ and $\Vsbar=-3\pm10$ \kms.
Fit A5 removed 15 sources found in Fit A1 to have a
motion component residual greater than $3\sigma$, used a Gaussian
probability distribution function for the residuals (ie, least-squares), 
and the same priors as A1.
Fits B1, C1 and D1 were similar to A5, except for the priors: B1 used
the solar motion priors of \citet{Schoenrich:10}
$(\U=11.1\pm2.0, \V=12.2\pm2.1, \W=7.2\pm2.0)$ \kms\ 
and no priors for source peculiar motions; C1 used
no solar motion priors and source peculiar motion priors
of $\Usbar=3\pm5$ and $\Vsbar=-3\pm5$ \kms; and 
D1 used flat priors for all parameters except
\Vo\ and \Vsbar, which were given unity probability between $\pm20$ \kms
of the initial Set-A values and zero probability outside this range.
The fit statistics listed are chi-squared ($\chisq$), the
number of degrees of freedom ($N_{dof}$), the number of sources used ($N_{sources}$),
and the Pearson product-moment correlation coefficient for parameters \Ro\ and \To\ ($r_{\Ro,\To}$). 
               }
\label{table:fits}
\end{deluxetable}

\subsection{Model A5}

With the resulting ``clean'' data set of 80 sources, we performed a 
least-squares fit (assuming a Gaussian PDF for the data uncertainties).  We
used the same loose priors (Set-A) as for model A1, namely solar motion
components $\U = 11.1\pm1.2$, $\V = 15\pm10$, $\W = 7.2\pm1.1$ \kms\ and 
average peculiar motions for HMSFRs of $\Usbar = 3\pm10$ and $\Vsbar =-3\pm10$ \kms.
This resulted in the parameter estimates listed under fit A5 in 
Table \ref{table:fits}.   This model produced a good $\chisq=224.9$ for 
232 degrees of freedom and estimates of $\Ro=8.34\pm0.16$ kpc and $\To=240\pm8$ \kms.
We find $\Tdot=-0.2\pm0.4$ \kmsperkpc, indicating a very flat rotation curve for 
the Milky Way between radii of $\approx5$ and 16 kpc from the Galactic center.

\begin{figure}[htp]

\epsscale{0.8} 
\plotone{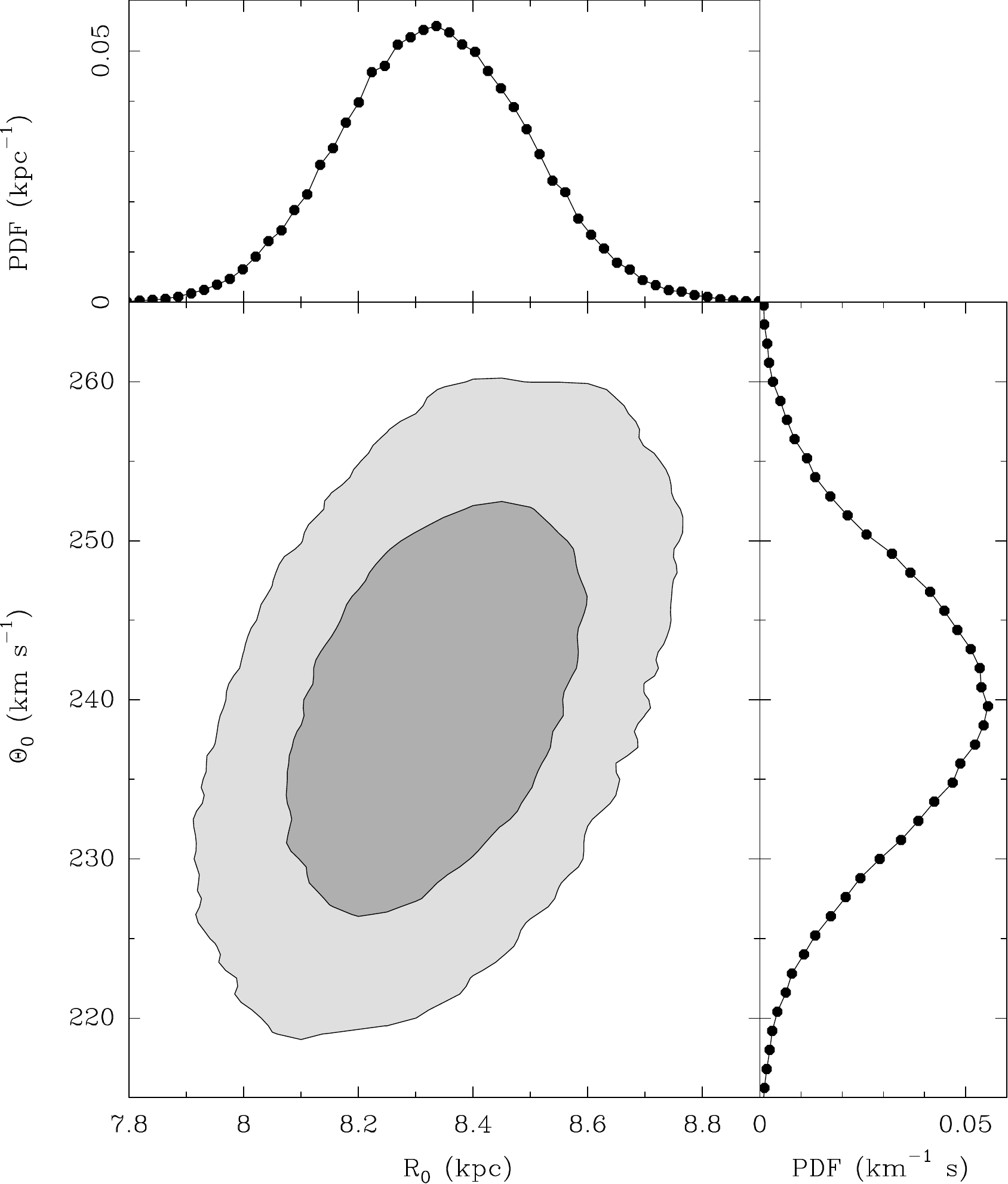}  
\caption{\small
Joint and marginalized posteriori probability density distributions for 
\Ro\ and \To\ for model fit A5.  Priors for the solar motion were
$\U = 11.1\pm1.2$, $\V = 15\pm10$, $\W = 7.2\pm1.1$ \kms\ and for the
average peculiar motions for HMSFRs were 
$\Usbar = 3\pm10$ and $\Vsbar =-3\pm10$ \kms.  Contours enclose 95\% and 68\%
probabilities.  The Pearson product-moment correlation coefficient for 
\Ro\ and \To\ is 0.46.
        }  
\label{fig:RoToPDF}
\end{figure}

Compared to the preliminary results of \citet{Reid:09b} based on 16 sources, where the
Pearson product-moment correlation coefficient for \Ro\ and \To\ was high, 
$r_{\Ro,\To}=0.87$, with the larger number of sources and a better distribution
across the Galaxy, these parameters are significantly less correlated, $r_{\Ro,\To}=0.46$.
The joint and marginalized PDFs for these fundamental Galactic parameters are 
displayed in Figure \ref{fig:RoToPDF}.

The circular velocity parameters are still correlated (see \S\ref{sect:correlations}), 
but linear combinations of these parameters are well determined: 
$\To+\V=255.2\pm5.1$ and $\V-\Vsbar=17.1\pm1.0$.
Also, the angular rotation rate for the Sun's orbit about the Galactic center 
is constrained to $\pm1.4$\% accuracy: 
$(\To+\V)/\Ro=30.57\pm0.43$ \kmskpc.   This value is consistent with the reflex of
the {\it apparent} motion of Sgr~A*, the assumed motionless supermassive black hole 
at the center of the Galaxy, which gives $30.26\pm0.12$ \kmskpc\ \citep{Reid:04}.

The component of solar motion in the direction of Galactic rotation, $\V$,
estimated to be $15.6\pm6.8$ \kms\ is better constrained than the prior of $15\pm10$ \kms.  
It is consistent with the {\it local} estimate (relative to Solar Neighborhood stars) 
of 12 \kms\ \citep{Schoenrich:10} and the {\it global} estimate of \citet{Bovy:12} of 
$26\pm3$ \kms\ (relative to stars across the Milky Way).

\subsection{Model B1}

In order to explore the sensitivity of the modeling to our priors,
we fit the clean data set using the Set-B priors: 
adopting the latest Hipparcos measurement of the solar motion of $\U = 11.1\pm1.2$, 
$\V = 12.2\pm2.1$, $\W = 7.2\pm1.1$ \kms\ \citep{Schoenrich:10} 
and no prior information on the average peculiar motion of the HMSFRs.  
This resulted in parameter estimates similar
to those of model A5, \eg\ $\Ro=8.33\pm0.16$ kpc and $\To=243\pm6$ \kms.
The quality of fit, as measured by $\chisq=225.1$ for 232 degrees of freedom,
was comparably good as for model A5.   The average velocity lag of the HMSFRs
relative to circular orbits, which was not constrained by priors, was 
$\Vsbar = -5.0 \pm 2.1$ \kms.  This is comparable to that found by \citet{Reid:09b},
after correcting for the 7 \kms\ difference in the adopted solar motion values. 

\subsection{Model C1}

Given the current uncertainty in the \V\ component of solar motion, we fit the data 
with the Set-C priors, assuming no prior information for the solar motion, but using 
a stronger prior than for model A5 for the average peculiar motion of the HMSFRs: 
$\Usbar = 3\pm5$ and $\Vsbar =-3\pm5$ \kms.  As for model B1, we found most 
parameter estimates to be similar to model A5, eg, $\Ro=8.30\pm0.19$ kpc and 
$\To=239\pm8$ \kms.  For the solar motion, we find $\U = 9.9\pm2.0$, 
$\V =14.6\pm5.0$, and $\W = 9.3\pm1.0$ \kms.   The \V\ value is consistent
with revised \citet[12 \kms]{Schoenrich:10} solar motion, but differs by
$2\sigma$ from the \citet{Bovy:12} estimate.

\subsection{Model D1}

In order to facilitate the use of the results presented here with other Galactic
parameter estimates, we perfomed a fit with essentially no informative priors.
We did this by taking the A5 (Set-A) initial parameter values and assuming
flat priors for all parameters except for \V\ and \Vsbar.  For these parameters
we assumed equal probability for values within $\pm20$ \kms\ of the initial values
and zero probability outside this range in order to exclude unreasonable parameter values.
The parameters that remain well determined include $\Ro=8.29\pm0.21$ kpc, 
$\To=238\pm15$ \kms, $\Tdot=-0.1\pm0.4$ \kmskpc, $\U=9.6\pm3.9$ \kms,
$\W=9.3\pm1.0$ \kms, and $\Usbar=1.6\pm3.9$ \kms.   The correlated velocity terms,
\V\ and \Vsbar\ displayed nearly flat posteriori PDFs over their allowed ranges.
However, linear combinations involving these parameters are very well constrained,
$\To+\V = 253.8\pm6.4$ \kms\ and $\V-\Vsbar = 17.2\pm1.2$ \kms,
as well as the angular rotation rate of the Sun about the Galactic center,
$(\To+\V)/\Ro=30.64\pm0.41$ \kmskpc.

\subsection{Rotation Curves} \label{sect:rotationcurves}

Next, we investigated the sensitivity of the fundamental
Galactic parameters, \Ro\ and \To, to alternative rotation curves.  
When fitting, we replaced the simple linear form, $\Theta(R)=\To+\Tdot~(R-\Ro)$,
with the empirically determined functions of $\Theta(R)$ of \citet{Clemens:85},
the power-law parameterization of \citet{Brand:93}, a polynomial, and 
the ``universal'' rotation curve of \citet{Persic:96}.  
We adopted the Set-A priors in order to facilitate comparisons with the A5 fit.  
Table~\ref{table:rotationcurves} presents the fitting results for these rotation 
curves.

\begin{deluxetable}{lccccc}
\tablecolumns{6} \tablewidth{0pc} 
\tablecaption{Rotation Curve Results}
\tablehead {
  \colhead{} &\colhead{C-10} &\colhead{C-8.5} &\colhead{BB} &\colhead{Poly} &\colhead{Univ}   
           }
\startdata
Parameter Estimates \\
\Ro~(kpc)    &$8.36\pm0.16$  &$8.12\pm0.14$ &$8.34\pm0.16$ &$8.34\pm0.17$  &$8.31\pm0.16$ \\
\To~(\kms)   &$237\pm8\q$    &$221\pm8\q$   &$240\pm9\q$   &$241\pm9\q$    &$241\pm8\q$   \\
\U~(\kms)    &$10.1\pm1.8~$  &$10.5\pm1.8\p$&$10.5\pm1.8\p$&$10.7\pm1.7\p$ &$10.5\pm1.7\p$\\
\V~(\kms)    &$19.4\pm6.8~$  &$25.0\pm6.8~$ &$15.5\pm6.8~$ &$14.7\pm6.8~$  &$14.4\pm6.8~$ \\
\W~(\kms)    &$8.9\pm1.0$    &$8.9\pm1.0$   &$8.8\pm1.0$   &$8.8\pm0.9$    &$8.9\pm0.9$   \\
\Usbar~(\kms)&$2.4\pm2.1$    &$2.6\pm2.0$   &$2.8\pm2.0$   &$2.8\pm2.0$    &$2.6\pm2.1$   \\
\Vsbar~(\kms)&$+3.4\pm6.8\p$ &$+8.5\pm6.8\p$&$-1.5\pm6.8$  &$-1.4\pm6.8\p$ &$-1.4\pm6.8\p$\\
\aone~(\kms) &...            &...           &$240\pm9~$    &$241\pm9~$     &$241\pm8~$    \\
\atwo        &...            &...           &$~0.00\pm0.02$&$~~0.5\pm3.7$  &$~0.90\pm0.06$\\
\athr        &...            &...           &...           &$-15.1\pm8.4~$ &$~1.46\pm0.16$\\
\\
Fit Statistics \\
$\chisq$     &229.7          & 248.1        & 225.2        & 221.9         &214.5         \\
$N_{dof}$    &233            & 233          & 231          & 230           &230           \\
$N_{sources}$&80             & 80           & 80           &  80           & 80           \\
$r_{\Ro,\To}$&0.46           &0.36          &0.48          &0.47           &0.47          \\
\enddata
\tablecomments{\footnotesize
Rotation curves C-10 and C-8.5 are from \citet{Clemens:85} for old and revised 
IAU recommended values of ($\Ro=10$ kpc, $\To=250$ \kms) and ($\Ro=8.5$ kpc, 
$\To=220$ \kms), respectively.  These curves have been scaled by the fitted values 
for these fundamental parameters.  Note the higher $\chisq$ value for the
C-8.5 model compared to the others in the table.
The ``BB'' rotation curve from \citet{Brand:93} is a power law in radius: 
$\Theta(R) = \aone (R/\Ro)^\atwo$.
The ``Poly'' model is second-order polynomial in radius:
$\Theta(R) = \aone + \atwo \rho + \athr \rho^2$, where $\rho=(R/\Ro)-1$.
The ``Univ'' curve is a universal rotation curve \citep{Persic:96}, where 
\aone\ is the rotation speed at the optical (83\% light) radius, and the other 
parameters are dimensionless and provide the shape.
For the latter three models, \To\ is not an independently adjustable parameter;
instead it is calculated from \aone, \atwo, and \athr.
               }
\label{table:rotationcurves}
\end{deluxetable}

\citet{Clemens:85} supplied two curves with different shapes: 
one assuming the old IAU constants (C-10) of $\Ro=10$~kpc 
and $\To=250$~\kms\ and the other assuming the revised constants (C-8.5) of $\Ro=8.5$~kpc 
and $\To=220$~\kms\ currently in widespread use.  
The C-10 model has rotational speeds that rise faster with radius than the 
C-8.5 model.   For either model, we fitted for different values of \Ro\ (which 
we used to scale model radii) and \To\ (which we used to scale rotation speeds).  

\citet{Brand:93} parameterize their rotation curve (BB) as a power law in Galactocentric
radius, $R$, with potentially three adjustable parameters:
$\Theta(R) = \aone (R/\Ro)^\atwo + \athr$.
For a flat rotation curve ($\atwo=0$), parameters \aone\ and \athr\
become degenerate.  Since the Galaxy's rotation curve is nearly flat over
the range of radii we sample (see, eg, model A5 above),
we held \athr\ at zero, solving only for \aone\ and \atwo.  
Indeed, we find the power law exponent, $\atwo=-0.01\pm0.01$, essentially flat.
For this formulation, $\aone=\To$, and in Table \ref{table:rotationcurves}
we copy \aone\ to \To\ to facilitate comparison with other models.
  
As an alternative to a power law rotation curve, we fitted a second-order 
polynomial (Poly) in $\rho=(R/\Ro)-1$: $\Theta(R) = \aone + \atwo \rho + \athr \rho^2$.
The model fit parameters for this form of a rotation curve are similar to those
from models C-10, BB and Univ.

The universal (Univ) rotation curve of \citet{Persic:96} includes terms for
an exponential disk and a halo.  It can have three adjustable parameters: 
\aone, the circular rotation speed at the radius enclosing 83\% of the 
optical light ($R_{opt}$); 
$\atwo = R_{opt} / \Ro$; and
\athr, a core-radius parameter for the halo contribution, 
nominally 1.5 for an $L^*$ galaxy.  With flat priors for the three rotation
curve parameters, the posteriori PDF for $\atwo$ was bimodal, with the dominant
peak at $\atwo=0.9$ and a second peak with 50\% of the primary's amplitude
at $\atwo=0.1$.  Since the secondary peak seems unlikely, we refit the data using 
a prior for $\atwo$ of $1.2\pm0.5$.  We then obtained
similar parameter values as other models (see Table \ref{table:rotationcurves}), 
with the three adjustable rotation curve parameters of 
$\aone = 241\pm8$ \kms, $\atwo = 0.90\pm0.06$, and $\athr = 1.46\pm0.16$.  

All but one of the rotation curve models lead to similar values for the
fundamental Galactic parameters \Ro\ and \To\ as our A5 fit. 
Only the Clemens ``$\Ro=8.5$ kpc; $\To=220$ \kms'' (C-8.5) rotation curve results in a 
marginally significant change in estimates of \Ro\ and \To.   However, this fit 
has a significantly poorer quality ($\chi^2=248.1$ for 233 degrees of freedom) than, 
for example, the A5 fit ($\chi^2=224.9$ for 232 degrees of freedom), and we
do not consider this model further.  We conclude that the fundamental Galactic 
parameters \Ro\ and \To\ are reasonably insensitive to a wide variety of rotation 
curve shapes. 

\begin{figure}
\epsscale{0.95} 
\plotone{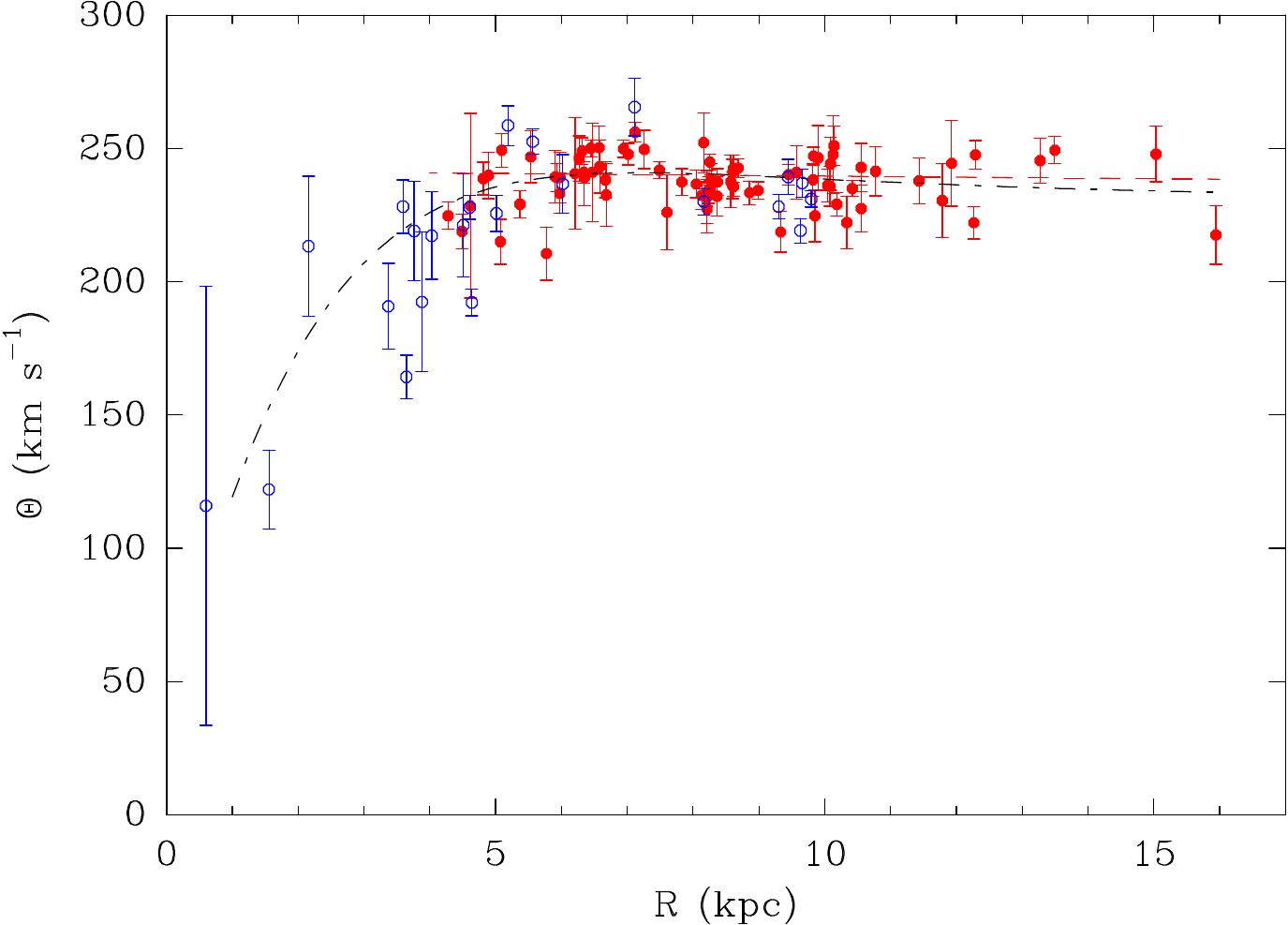}
\caption{\small
Rotation curve for all high mass star forming regions with measured parallax
and proper motion in Table \ref{table:parallaxes}.  Plotted is the 
circular velocity component, $\Theta$, as a function of Galactocentric radius, $R$.
The transformation from heliocentric to Galactocentric frames uses
the parameter values of fit A5, based only on sources with $R>4.0$ kpc;   
these sources are plotted with {\it filled red symbols}.
The sources not used in the final fitting are plotted with {\it open blue symbols}.
The {\it dashed red line} indicates the fitted rotation curve (model A5) given 
by $\Theta=\To -0.2(R-\Ro)$ \kms, where $R$ and $\Ro$ are in kpc.
The {\it dash-dot black line} is the best fit ``Universal'' rotation curve (model D1)
for spiral galaxies \citep{Persic:96}, which begins to capture the clear
velocity turn down for stars with $R\lax5.0$ kpc.
        }
\label{fig:rotationcurve}
\end{figure}

With full 3-dimensional location and velocity information, we can transform our 
heliocentric velocities to a Galactocentric reference frame and calculate the 
tangential (circular) speed for each HMSFR. Figure \ref{fig:rotationcurve} plots 
these speeds for {\it all} sources in Table \ref{table:parallaxes}.  
Most published rotation curves for the Milky Way 
have come from only one component of velocity (radial), often
using kinematic distances and {\it assuming} a value for \To.  
As such, the data in Fig \ref{fig:rotationcurve} represent a 
considerable advance.   See also the analysis of this data set by 
\citet{Xin:13}.  

It is important to remember that the
transformation from heliocentric to Galactocentric frames requires 
accurate values of \Ro, \U, and, most importantly, $\To+\V$, since the motion 
of the Sun has (by definition) been subtracted in the heliocentric frame. 
For most sources, increasing or decreasing the assumed value of $\To+\V$
would, correspondingly, move each data point up or down by about the same amount. 
Thus, the level of this, and essentially all published, rotation curves is
determined mostly by $\To+\V$.  Our results are the first to use fully 
3-dimensional data to strongly constrain all three parameters:
\Ro, \U\ and $\To+V$.  

The dashed line in Fig \ref{fig:rotationcurve} represent the linear rotation curve 
from the A5 fit, based only on sources with $R>4$ kpc.  Sources used in the
fit are plotted with filled symbols and the sources not used with open symbols.   
The dashed line indicates the expected rotation for sources in circular 
Galactic orbit (\ie\ $\Usbar=\Vsbar=0$).  There are now sufficient data to clearly
indicate that the rotation curve drops at Galactocentric radii $\lax4$ kpc.
However, given the likelihood for a significant non-axisymmetric gravitational
potential within $\approx4$ kpc of the center, more measurements are needed
before extending a rotation curve to this region as azimuthal terms may be needed. 

\subsection{Peculiar Motions of HMSFRs}

\begin{figure}
\epsscale{0.95} 
\plotone{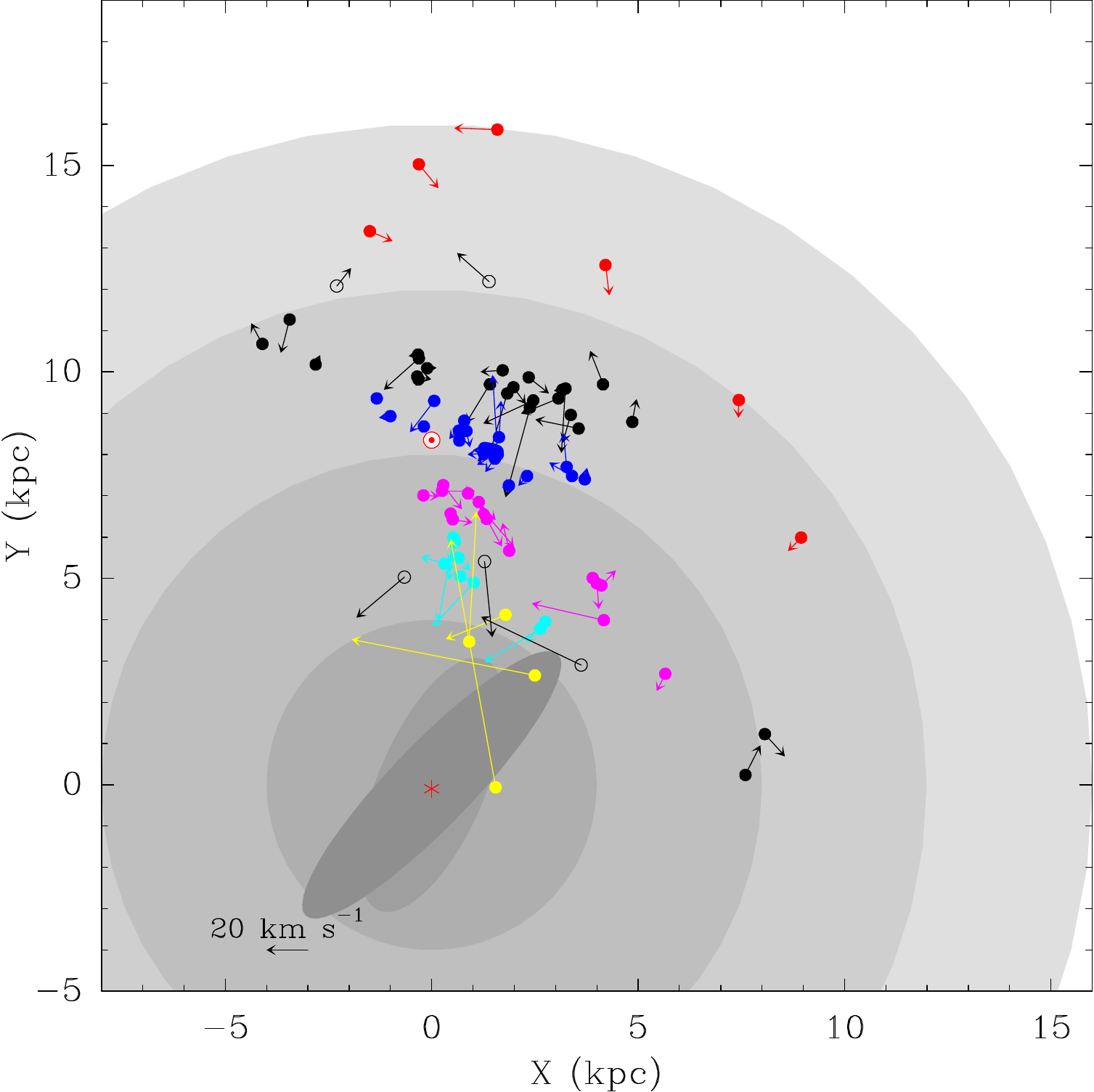}
\caption{\small
Peculiar (non-circular) motions of HMSFRs projected on the Galactic plane.
These motions ({\it arrows}) were calculated with parameter values from the A5 fit, specifically
$\Ro=8.34$ kpc, $\To=240$ \kms, $\Tdot=-0.2$ \kmsperkpc, and $\U=10.7$, $\V=15.6$ \kms\ 
(but without correction for \Usbar\ or \Vsbar).
Only sources with motion uncertainties $<20$ \kms\ are plotted.  A 20 \kms\ scale
vector is shown at the bottom left.  Spiral arm sources are color coded as describe
in Figure \ref{fig:parallaxes}.  The Galaxy rotates clockwise on this view from the NGP.
        }
\label{fig:peculiarmotions}
\end{figure}

Figure \ref{fig:peculiarmotions} shows the peculiar (non-circular) motions
of all sources in Table \ref{table:parallaxes} with motion uncertainties less 
than 20 \kms.  Similar results were described in the primary papers
presenting the parallaxes and proper motions for each arm 
\citep{Sato:13,Wu:13,Xu:13,Choi:13,Zhang:13,Hachi:13}.  For uniformity, 
here the motions were calculated using the A5 fit parameters (see Table \ref{table:fits}), 
but with zero correction for the average source peculiar motions.    
Typical peculiar motions are $\approx10$ \kms, but some sources have much
larger values.  For example, many sources in the Perseus arm in the 
Galactic longitude range $\approx100^\circ$ to $\approx135^\circ$ display peculiar
motions $\gax20$ \kms.  Many sources within $\approx4$ kpc of the Galactic Center
display even larger peculiar motions, probably indicating that the 
rotation curve used here is inadequate to describe their Galactic orbits,
especially in the presence of the Galactic bar(s).

\subsection{Parameter Correlations} \label{sect:correlations}

\begin{deluxetable}{lrrrrrrrr}
\tablecolumns{9} 
\tablewidth{0pc} 
\tablecaption{Parameter Correlation Coefficients}
\tablehead {
  \colhead{}  
  &\colhead{\Ro} &\colhead{$\To$} &\colhead{$\Tdot$} &\colhead{$\U$} &\colhead{$\V$} 
  &\colhead{$\W$} &\colhead{$\Usbar$} &\colhead{$\Vsbar$} 
            }
\startdata
  $\Ro$     &1.000   &0.465   &0.103   &0.452   &0.023   &$-$0.003&0.517    &$-$0.002 \cr
  $\To$     &0.465   &1.000   &0.136   &0.243   &$-$0.796&$-$0.009&0.171    &$-$0.809 \cr
  $\Tdot$   &0.103   &0.136   &1.000   &$-$0.124&$-$0.009&0.025   &$-$0.094 &$-$0.018 \cr
  $\U$      &0.452   &0.243   &$-$0.124&1.000   &$-$0.014&$-$0.017&0.839    &0.025    \cr
  $\V$      &0.023   &$-$0.796&$-$0.009&$-$0.014&1.000   &0.011   &$-$0.006 &0.990    \cr
  $\W$      &$-$0.003&$-$0.009&0.025   &$-$0.017&0.011   &1.000   &$-$0.002 &0.010    \cr
 $\Usbar$   &0.517   &0.171   &$-$0.094&0.839   &$-$0.006&$-$0.002&1.000    &0.028    \cr
 $\Vsbar$   &$-$0.002&$-$0.809&$-$0.018&0.025    &0.990   &0.010   &0.028    &1.000    \cr
\enddata
\tablecomments {\footnotesize
   Pearson product-moment correlation coefficients for the A5 fit 
   calculated from $10^6$ McMC trial parameter values thinned by a factor of 10.  
   Parameter definitions are given in the text and the notes in 
   Table~\ref{table:model}. 
              }
\label{table:correlations}
\end{deluxetable}

\begin{figure}[ht]
\epsscale{0.7} 
\plotone{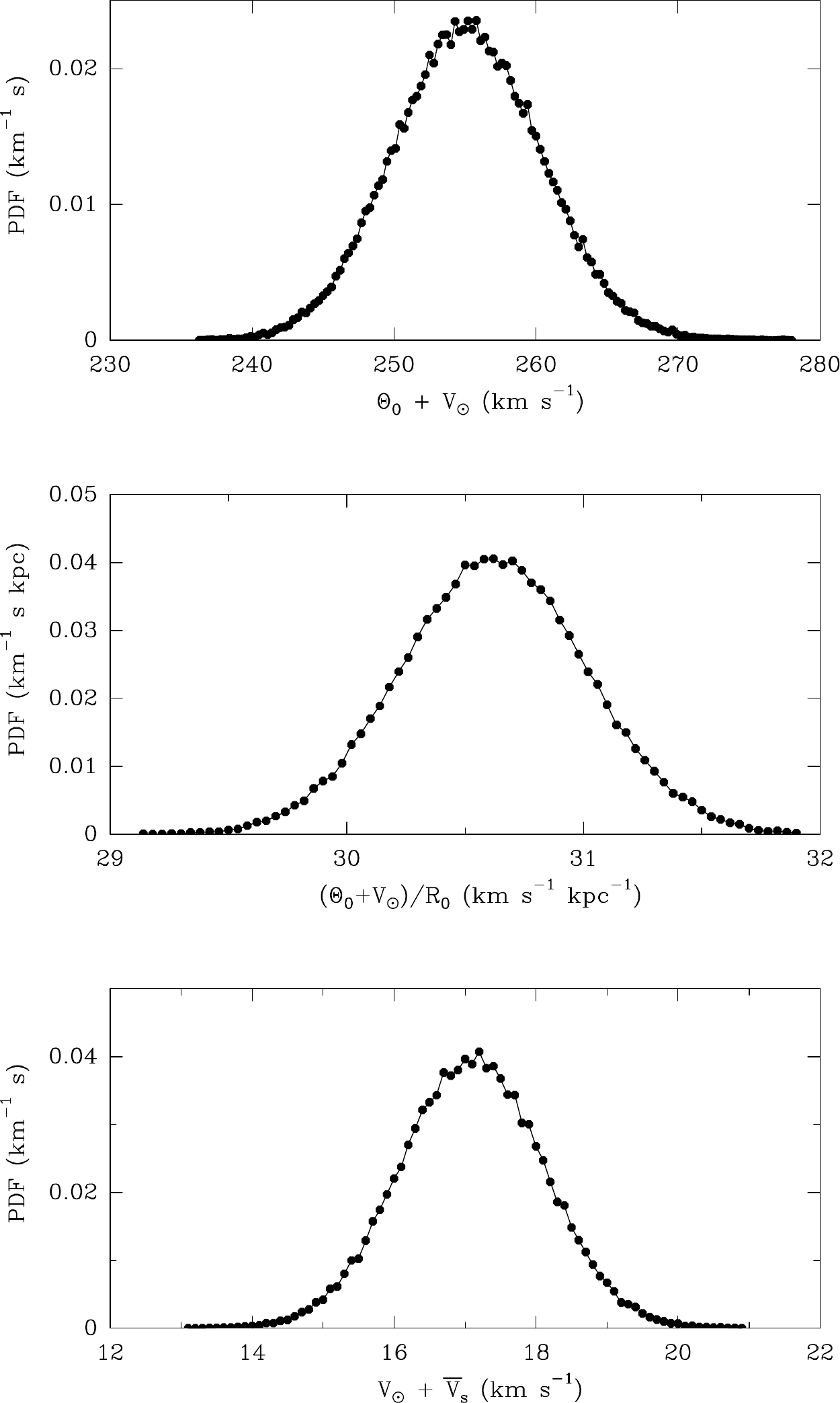}
\caption{\small
Marginalized poisteriori probability density distributions for 
correlated circular velocity parameters from fit A5. 
{\it Top panel:} the circular orbital speed of
the Sun: $\To + \V$.  {\it Middle panel:} the angular orbital speed of
the Sun: $(\To+\V)/\Ro$. {\it Bottom panel:} difference between the circular
solar and average source peculiar motions: $\V - \Vsbar$. 
        }
\label{fig:3PDFs}
\end{figure}

The Pearson product-moment correlation coefficients, $r$, for all parameters 
from fit A5 are listed in Table \ref{table:correlations}. 
In the preliminary analysis of 16 HMSFRs with parallaxes and proper motions 
of \citet{Reid:09b}, the estimates of \Ro\ and \To\ were strongly correlated 
($\rPpm=0.87$).  However, with the much larger set of HMSFRs that 
covers a larger portion of the Galaxy, the correlation between \Ro\ and \To\ 
estimates is now moderate: $\rPpm=0.465$ for our reference A5 fit.   
However, there remains a significant anti-correlation between \To\ and \V\ 
($r_{\To,\V}=-0.809$), as well as a strong correlation between \V\ and \Vsbar\ 
($r_{\V,\Vsbar}=0.990$).
As suggested by the fitted parameter values in Table \ref{table:fits}, 
our data strongly constrain the following combinations of these correlated parameters:
$\To + \V=255.2\pm5.1$ \kms\ and $\V - \Vsbar=17.1\pm1.0$ \kms.
Also, the combination of parameters that yield the angular orbital speed of the Sun 
about the Galactic center, $(\To+\V)/\Ro = 30.57 \pm 0.43$ \kmskpc, is more
tightly constained than the individual parameters.
Figure \ref{fig:3PDFs} shows the marginalized PDFs for these combinations of
parameters.

\subsection{Comparison with Other Modeling Approaches}

Other groups have analyzed parallax and proper motion data sets
from the BeSSeL Survey and the VERA project, focusing on 
different assumptions and results.  \citet{Bovy:09} confirmed the counter rotation 
of HMSFRs (assuming $\V=5$ \kms) noted by \citet{Reid:09b} 
and argued for a comparable value for \To\ $(246\pm30$ \kms), but with 
considerably lower significance.  
Alternatively, \citet{McMillan:10} found that the \V-component of solar motion 
of 5 \kms, provided by \citet{Dehnen:98}, should be raised to $\approx12$ \kms, 
thereby reducing the estimated counter-rotation of HMSFRs. 
\citet{Bobylev:10}, using 28 parallaxes available at that time and
a Fourier analysis technique, estimated $\To=248\pm14$ \kms\ and $\V=11.0\pm1.7$ \kms,
assuming $\Ro\equiv8.0$ kpc.  Finally, \citet{Honma:12}, using 52 parallaxes,
including some low-mass star forming regions, estimated $\Ro=8.05\pm0.45$ kpc
and $\To=238\pm14$ \kms\ for $\V\equiv12$ \kms.

The \citet{Bovy:09} re-analysis of our preliminary data employed a different approach 
than that of \citet{Reid:09b}.  Bovy \etal\ treat 
the elements of the velocity dispersion tensor of the HMSFRs as free parameters.  
These parameters give the expected deviations (variances and covariances) of the 
velocity data from a smooth, axi-symmetric model of Galactic rotation and are used 
to adjust the weights applied to the different velocity components when fitting the data.  
However, while Bovy \etal\ found a significant trace for the tensor, the velocity 
dispersion parameters were only marginally constrained; formally none of the diagonal 
components had $>2.8\sigma$ formal significance.  
Also, their values for the radial and tangential components were nearly identical, 
suggesting that little is gained by making these free parameters versus adopting a single 
physically motivated value ($\sigma_{Vir}$) as we have done.  Note that our value for 
$\sigma_{Vir}$ is comparable to the dispersion parameter ($\Delta_v$) values found by 
\citet{McMillan:10}, which range from about 6 to 10 \kms, but is considerably smaller 
than those of \citet{Bovy:09} of $\approx20$ \kms.  The reason for this difference 
is unclear, but might reflect different treatments of outlying data and/or increased
parameter correlations associated with the 6 extra parameters used in solving 
for the tensor elements.

\section{Discussion }   \label{sect:discussion}

\subsection{Solar Motion} \label{sect:solar_motion}

If one adopts the theoretically motivated prior that HMSFRs have small peculiar 
motions (Set-C with no prior on the solar motion), then model fit C1 indicates 
$\V=14.6\pm5.0$ \kms.  This is a {\it global} measure of the peculiar motion of 
the Sun and, as such, is relative to a ``rotational standard of rest'' as opposed 
to a Local Standard of Rest (LSR), defined relative to stellar motions in the 
Solar Neighborhood.  If Solar Neighborhood stars (extrapolated to a zero-dispersion 
sample) are, on average, stationary with respect to a circular orbit, then these two 
solar motion systems will be the same.  Our estimate of $\V$ is consistent with the
$12$ \kms\ value of \citet{Schoenrich:10}, measured  with respect to Solar neighborhood 
stars, but there is some tension between our global estimate of \V\ and that of
\citet{Bovy:12} of $26\pm3$ \kms, as these two estimates differ by about $2\sigma$.
However, if one drops the prior that HMSFRs have small peculiar motions, then
our result loses significance.   

The large counter-rotation of HMSFRs, originally suggested by \citet{Reid:09b}, 
was based on the initial Hipparcos result of \citet{Dehnen:98} that $\V=5$ \kms.  
As the outcome of the \citet{Schoenrich:10} re-analysis of Hipparcos data, which gives 
$\V=12$ \kms, supersedes that lower $\V$ value, it now appears that any average 
counter-rotation of HMSFRs is $\lax5$ \kms.  Given that we strongly constrain 
$\V-\Vsbar=17.1\pm1.0$ \kms, were one to independently constrain $\V$ with $\pm2$ \kms\ 
accuracy, the issue of HMSFR counter rotation could be clarified.

While our estimate of \V\ has a large uncertainty (owing to correlations with \To\ and 
\Vsbar), we find \U\ and \W\ are well constrained.  In fit D1, in which no informative
prior was used for the components of motion either toward the Galactic center or 
perpendicular to the Galactic plane, we find that $\U=9.6\pm3.9$ and $\W=9.3\pm1.0$ \kms,
respectively.  Our estimate of the Sun's motion toward the Galactic center is in
agreement with most other estimates, \eg\ $11.1\pm1.2$ \kms\ by \citet{Schoenrich:10} 
and $10\pm1$ \kms\ by \citet{Bovy:12}; see also the compilation of estimates 
by \citet{Coskunoglu:11}.   

The solar motion component perpendicular to the Galactic plane, $\W$, is generally
considered to be straight forwardly determined and recent estimates typically range 
between 7.2 \kms\ \citep{Schoenrich:10} (relative to local stars within $\sim0.2$ kpc)
and 7.6 \kms\ \citep{Feast:97} (relative to stars within $\sim3$ kpc), 
with uncertainties of about $\pm0.5$ \kms.  We find a slightly larger value of 
$\W=9.3\pm1.0$ \kms\ (for model D1 which used no informative priors for the solar motion), 
which may be significant; the difference between the locally and our globally measured 
value (\ie\ relative to stars across the Galaxy) is $2.1\pm1.1$ \kms.  
Note that one might expect a small difference between measurements with respect to 
a local and a global distribution of stars were the disk of the Galaxy to 
precess owing to Local Group torques.   Simulations of galaxy interactions in a group 
suggest that a disk galaxy can complete one precession cycle over a Hubble time.
Were the Milky Way to do this, one would expect a vertical pecessional motion 
at a Galactocentric radius of the solar neighborhood of order $\Ro\Ho~\sim0.6$ \kms. 
It is possible that the differences in the local and global estimates of $\W$
can, in part, be explained in this manner. 

\subsection{Galactic Rotation Curve and Disk Scale Length}

Among the various forms of rotation curves that we fit to the data, 
the universal curve advocated by \citet{Persic:96} to apply to most
spiral galaxies yielded the best fit (see discussion in \S\ref{sect:rotationcurves},
Table \ref{table:rotationcurves} and Fig.~\ref{fig:rotationcurve}).  This rotation curve 
matches the flat to slightly declining run of velocity with Galactocentric radius from 
$R\approx5\rightarrow16$ kpc, as well as reasonably tracing the decline
in orbital velocity for $R\lax5$ kpc.   However, many of the sources
near the Galactic bar(s) cannot be well modeled with any axi-symmetric
rotation curve.

The best fit value for our $\atwo$ parameter ($R_{opt}/\Ro$), coupled with our estimate 
of $\Ro=8.34\pm0.16$, locates $R_{opt}$ at $7.5\pm0.52$ kpc.  
The $\atwo$ parameter is sensitive to the slope of the rotation curve (near $R_{opt}$)
and the radius at which it turns down toward the Galactic center.  For example, setting 
$\atwo=0.7$ steepens the rotation curve at large radii and moves the turn down radius to 
$\approx3.5$ kpc, while setting $\atwo=1.1$ flattens the rotation curve and increases the 
turn down radius to $\approx6.5$ kpc.  Given that the (thin) disk 
scale length, $R_D = R_{opt} / 3.2$ \citep{Persic:96}, we estimate $R_D=2.44\pm0.16$ kpc.  
Estimates of $R_D$ in the literature range from $\approx 1 \rightarrow 6$ kpc 
\citep{Kent:91,Chang:11,McMillan:11}, with most consistent with a value between 
$2\rightarrow3$ kpc.  Our estimate is also consistent with that of \citet{Porcel:98}
who modeled the positions and magnitudes of 700,000 stars in the Two Micron Galactic
Survey database and found $R_D=2.3\pm0.3$ kpc and, more recently, \citet{Bovy:13}, 
who modeled the dynamics of $\approx16,000$ stars from the SEGUE survey and concluded 
that $R_D=2.14\pm0.14$ kpc.

\subsection{The Distance to the Galactic Center: \Ro}

Models A5, B1 and C1, which used different combinations of solar motion
and/or average source peculiar motion priors, have comparable $\chi^2$ values
and all parameter estimates are statistically consistent.   
Because the priors for Model A5 are the least restrictive in keeping with
current knowledge, we adopt those parameters as representative.  Specifically,
we find $\Ro=8.34\pm0.16$ kpc, $\To=240\pm8$ \kms\ and $\Tdot=-0.2\pm0.4$ \kmsperkpc.
As noted in \S\ref{sect:correlations} and \S\ref{sect:rotationcurves}, with the much 
larger data set now available, estimates of \Ro\ and \To\ are no longer strongly 
correlated and appear fairly insensitive to the assumed nature of the rotation curve.
These parameter estimates are consistent with, but significantly better than, 
the preliminary values of $\Ro=8.4\pm0.6$ kpc, $\To=254\pm16$ \kms\ and a nearly flat 
rotation curve reported in \citet{Reid:09b}, based on parallaxes and proper 
motions of 16 HMSFRs and assuming $\V=5$ \kms, and $\Ro=8.05\pm0.45$ kpc and 
$\To=238\pm14$ \kms\ from \citet{Honma:12}, based on a sample of 52 sources and 
assuming $\V=12$ \kms.    

While there are numerous estimates of the distance to the Galactic center in
the literature (\eg\ \citet{Reid:93}), here we only compare those based on
direct distance measurements.  A parallax for the water masers in Sgr~B2,
a star forming region projected less than 0.1 kpc from the Galactic center,
indicates $\Ro=7.9\pm0.8$ kpc \citep{Reid:09c}, consistent with, but 
considerably less accurate than, our current result.
More competitive estimates of \Ro\ come from the orbits of ``S-stars'' about
the supermassive blackhole Sgr A*.  Combining the nearly two decades of data
from the ESO NTT/VLT \citep{Gillessen:09a} and Keck \citep{Ghez:08} telescopes that trace 
more than one full orbit for the star S2 (a.k.a. S0-2), \citet{Gillessen:09b}
conclude that $\Ro=8.28\pm0.33$ kpc.  Recently the Keck group, extending 
their time sequence of observations by only a few years, announced a value of
$\Ro=7.7\pm0.4$ kpc \citep{Morris:12}, in mild tension both with the
\citet{Gillessen:09b} analysis and our parallax-based result.  However,
in the latest publication of the Keck group, \citet{Do:13} combined modeling of
the distribution and space velocities of stars within the central 0.5 pc of the Galactic 
center with the stellar orbital result for star S0-2 \citep{Ghez:08} and conclude 
that $\Ro=8.46^{+0.42}_{-0.38}$ kpc, removing any tension with our estimate and 
that of the ESO group.  We conclude that our estimate of $\Ro=8.34\pm0.16$ kpc is 
consistent with that from the Galactic center stellar orbits and is likely the 
most accurate to date.

\subsection{The Circular Rotation Speed at the Sun: \To}

Over the last four decades there have been many estimates of \To\ ranging
from $\sim170 \rightarrow 270$ \kms\ \citep{Kerr:86,Olling:98}.  
Focussing the discussion to the more direct measurements, two recent
studies favor a lower and one a higher value of \To\ than our  
estimate of $\To=240\pm8$ \kms.
\citet{Koposov:10} model the orbit of the GD-1 stream from a tidally disrupted 
stellar cluster in the Milky Way halo and estimate $\To+\V=221\pm18$, where the 
\citet{Dehnen:98} solar motion component of $\V=5$ \kms\ was adopted.
Recently, \citet{Bovy:12} modeled line-of-sight velocities of 3365 stars
from APOGEE and find $\To=218\pm6$ \kms, but with a large value for the
solar motion component in the direction of Galactic rotation, $\V=26\pm3$ \kms.
Their full tangential speed $\To+\V=242^{+10}_{-3}$ is consistent with our value
of $252.2\pm4.8$ \kms, suggesting the discrepancy between the Bovy \etal\ 
and our results are probably caused by differences in the solar motion.
However, another recent study by \citet{Carlin:12}, modeling the Sagittarius
tidal stream, yields $\To$ estimates from $232 \rightarrow 264$ \kms.

Our data also strongly constrain the angular rotation of the Sun about the 
Galactic center, $(\To+\V)/\Ro=30.57\pm0.43$ \kmskpc.
This value can be compared with an independent and direct estimate based on
the proper motion of Sgr~A*, interpreted as the reflex motion from the 
Sun's Galactic orbit, of $30.24\pm0.12$ \kmskpc\ \citep{Reid:04}.
For $\Ro=8.34\pm0.16$ kpc, the proper motion of Sgr A* translates to 
$\To+\V=252.2\pm4.8$ \kms, in good agreement with the parallax results.
We conclude that $\To$ exceeds the IAU recommended value of 220 \kms\ 
with $>95$\% probability provided that $\V\lax23$ \kms.  Clearly, independent {\it global}
measures of $\V$ are critical to establish $\To$ and $\Vsbar$ with high accuracy.

Changing the value of \To\ would have widespread impact in astrophysics.
For example, increasing \To\ by 20 \kms\ with respect to the IAU recommended 
value of 220 \kms\ reduces kinematic distances by about 10\%, leading to a decrease 
of 20\% in estimated young star luminosities, a corresponding decrease in estimated
cloud masses, and a change in young stellar object ages.  
Estimates of the total mass of the dark matter halo of the Milky Way
scale as $V_{max}^2~R_{Vir}$.  Since the maximum in the rotation curve ($V_{max}$) 
and the Virial radius ($R_{Vir}$) scale linearly with \To, the mass of the halo 
scales as $\Theta_0^3$, leading to a 30\% increase in the estimate of the 
Milky Way's (dark-matter dominated) mass.
This, in turn, affects the expected dark-matter annihilation signal \citep{Finkbeiner:09},
increases the ``missing satellite'' problem \citep{Wang:12}, and increases the
likelihood that the Magellanic Clouds are bound to the Milky Way \citep{Shattow:09}.

\subsection{The Hulse-Taylor Binary Pulsar and Gravitation Radiation}

An interesting example of the effects of Galactic parameters on fundamental
physics comes from the Hulse-Taylor binary pulsar.  The dominant uncertainty in 
measuring the gravitational radiation damping of the binary's orbit 
comes from the need to correct for the effects of the Galactic accelerations of 
the Sun and the binary \citep{Damour:91,Weisberg:10}.  
These accelerations contribute $\approx1$\% to the {\it apparent} orbital period decay.  
In 1993 when the Nobel Prize was awarded in part for this work, the IAU recommended 
values were $\Ro=8.5\pm1.1$ kpc and $\To=220\pm20$ \kms\ \citep{Kerr:86}.  
Using these Galactic parameters, the formalism of Damour \& Taylor, improved pulsar 
timing data of Weisberg, Nice, \& Taylor, and a pulsar distance of 9.9~kpc, 
the binary's orbital period decays at a rate of $0.9994\pm0.0023$ times that 
prediction from general relativity (GR).  
Using the improved Galactic parameters from the A5 fit ($\Ro=8.34\pm0.16$ kpc and 
$\To=240\pm8$ \kms), gives a GR test value of $0.9976\pm0.0008$.  
This provides a three-fold improvement in accuracy. Both of these examples assumed a 
distance to the binary pulsar of 9.9~kpc \citep{Weisberg:08}.  Given the improvement
in the Galactic parameter values, the dominant uncertainty in the GR test now
is the uncertain pulsar distance.  A pulsar distance of 7.2 kpc would bring the
GR test value to 1.0000 and a trigonometric parallax accurate to $\pm8$\%, 
which is possible with in-beam calibration with the VLBA, would bring the contribution 
of distance uncertainty down to that of the current Galactic parameter uncertainty.
Alternatively, if one assumes GR is correct, the current improvement in Galactic
parameters suggests that the Hulse-Taylor binary pulsar's distance is $7.2\pm0.5$ kpc.

\vskip 0.5truein 
This work was partially funded by the ERC Advanced Investigator Grant GLOSTAR (247078).
The work was supported in part by the National Science Foundation of China
(under grants 10921063, 11073046, 11073054 and 11133008) and the Key
Laboratory for Radio Astronomy, Chinese Academy of Sciences.
AB acknowledges support by the National Science Centre Poland through grant 
2011/03/B/ST9/00627.

\vskip 0.5truein 
{\it Facilities:}  \facility{VLBA}, \facility{VERA}, \facility{EVN}

\end{document}